# The Higgs Field Governs the Interior Spacetime of Black Holes

Itzhak Bars[*]

*Department of Physics and Astronomy,*

*University of Southern California, Los Angeles, CA 90089*

## Abstract

The Higgs field is conventionally treated as a small local perturbation atop a large, constant vacuum value that uniformly permeates the universe. I propose instead that in regions of extreme gravitational intensity—such as near gravitational singularities—the Higgs field behaves in a profoundly non-perturbative manner. In such environments, spacetime and the Higgs field engage in a dynamic interplay that extends spacetime beyond the singularity, achieving geodesic completeness. The continuation beyond the singularity is dominated by antigravity effects, reshaping the causal structure of spacetime and enabling novel flows of matter and information, including traversal through singularities. In its standard form, the combined framework of the Standard Model (SM) and General Relativity (GR), as well as most of its extensions, fails to capture these phenomena due to its geodesic incompleteness. By contrast, a refined, locally conformal-symmetric formulation—denoted i(SM+GR)—naturally incorporates these effects. GR is not an optional component of i(SM+GR) but an essential ingredient. This framework preserves the empirical success of SM+GR in the low-energy regime while predicting striking new phenomena in extreme gravitational settings, including within black holes (on both sides of the singularity) and in pre–Big Bang cosmology. At the classical field theory level, i(SM+GR) offers fresh perspectives on the black hole information puzzle and provides a platform for locally scale-invariant generalizations, such as geodesically complete quantum field theory, string theory, and unified models of fundamental interactions. This paper presents the detailed derivations and explanations that underlie a condensed letter version recently published in [1].

---

[*]Electronic address: bars@usc.edu

Contents

## I. THE CASE FOR A GEODESICALLY COMPLETE GEOMETRY

The literature on black holes and cosmology has largely overlooked that spacetimes containing singularities are *geodesically incomplete.* This neglect likely stems from the expectation that quantum effects would drastically modify gravity near singularities and thereby repair the incompleteness. Yet, most tools developed to explore quantum gravity—including string theory—are themselves formulated on geodesically incomplete background fields when they address cosmological or black hole spacetimes, leaving them as incomplete as their field theory counterparts.

Some non-geometric approaches in string theory, such as matrix theory [2], attempt to address black hole physics through conformal field theory techniques. While these toy models may eventually shed light on the problem, they generally don't address the issues of singularities. Thus, geodesic incompleteness remains pervasive across traditional classical and quantum gravity frameworks. The result is the omission of potentially crucial physical effects, comparable in importance to the quantum corrections these theories aim to capture. A prime example is the black hole information puzzle, which could benefit from a new understanding of information flow in a geodesically complete and causal spacetime, as developed in this paper.

Resolving geodesic incompleteness must therefore be a *foundational step*, first at the classical level and only then within quantum theory. A geodesically complete framework for the Standard Model (SM) coupled to General Relativity (GR) naturally emerged from 2T-Physics in 4+2 dimensions and was later recast directly in 3+1 dimensions [3]. The essential new ingredient, from the 3+1 viewpoint, is a specialized form of local scale (Weyl) invariance[1]. Derived from 2T-physics, this symmetry enhances the behavior of SM+GR fields in strong gravity while preserving the empirically verified accuracy of SM+GR in weak gravity. The resulting theory, often denoted i(SM+GR)(the "i" signifying improvement

---

[1] The local scale transformation in i(SM+GR), derived from 2T-physics in 4+2 dimensions, corresponds to a specific class of general coordinate transformations that mix the extra (1+1) dimensions with the familiar 3+1 dimensions. While the resulting local scale transformation rules for fields share the same features with Weyl's original transformations, their conceptual origin is fundamentally different. In particular, i(SM+GR) does *not* introduce a Weyl vector as an independent gauge field degree of freedom. Reference [4] provides a concise summary of the 2T-physics foundations of i(SM+GR) in its Section III and Appendix. It also compiles references to works, following [3], that highlight the role of local conformal symmetry in i(SM+GR), including its applications to cosmology and black holes.

through local conformal symmetry), forms the backbone of the present work.

This enhanced theory closely mirrors the conventional SM+GR in accurately describing physical phenomena at low and well beyond current accelerator energies, but also addresses geodesic incompleteness by extending geodesics and field configurations to regions near and beyond black hole and cosmological singularities. Here, the local scale symmetry plays a pivotal role in establishing continuity through the singularity, offering profound new insights into the behavior of matter and energy under gravitationally extreme conditions. In these spacetime regions, where strong gravity and high energies dominate, new physical phenomena are predicted within classical i(SM+GR). Such physical effects, that are likely to survive in their quantum versions, have indirect consequences that could be observed outside of black holes in the regions of spacetime we inhabit.

The complete geometry revealed by i(SM+GR) includes not only the familiar gravity domains but also new *antigravity patches* that lie beyond every black hole and cosmological singularities. In these regions, gravity becomes repulsive. The gravity together with the antigravity domains yield a fully geodesically complete spacetime that encompasses both sides of every black hole singularity and every Big Bang/Big Crunch singularity.

The root cause of geodesic incompleteness is the absence of the antigravity domains in the conventional SM+GR or in geodesically incomplete quantum gravity models. So, the novel physics predicted by i(SM+GR) cannot be obtained from the traditional approach, that typically assumes a breakdown of physics at singularities, and defers the resolution to an as-yet-unsettled quantum gravity model. This breakdown assumption has proven unfruitful. In contrast, i(SM+GR), guided by local conformal symmetry, offers a *complete spacetime structure* and the tools to explore new phenomena while maintaining the successes of SM+GR at low energies. Moreover, a proposal for how to extend the i(SM+GR) type of conformal symmetry to string theory has been made in [5]. This development may represent a decisive step toward a unified quantum gravity theory.

Thus, i(SM+GR) stands out as a well motivated coherent framework that: matches SM+GR in all experimentally verified domains, extends geodesics beyond singularities in a complete spacetime that unifies gravity with antigravity, predicts new physics in regions of extreme gravity near singularities, and makes new predictions outside black holes while preserving consistency with low-energy observations.

The remainder of this paper is organized as follows. Section II reviews the action of

i(SM+GR), highlighting its essential features, phenomenological accuracy, and equations of motion. Section IV solves the equations in asymptotic regimes to determine the vacuum state, describes the complete geometry of gravity and antigravity domains, and exhibits continuous geodesics that traverse black hole singularities. Section V constructs the extended Kruskal–Szekeres $(u, v)$ unified spacetime for the new AdSSdS black hole, combining an asymptotically dS gravity domain , with an asymptotically AdS antigravity domain. Section VI presents the corresponding Kruskal–Szekeres and Penrose diagrams of this geodesically complete black hole and discusses their geometric and causal properties. Finally, Section VII considers the global structure of the complete universe, including implications for black hole quantum information and unitarity.

## II. THE ACTION FOR I(SM+GR) AND ITS NOTABLE FEATURES

In this section, I analyze the vacuum state of i(SM+GR) with geodesically complete geometry, emphasizing its profound implications for the interior of black holes. Due to space limitations, I will not repeat the detailed discussion of the virtues of i(SM+GR) presented in [3][4]. Instead, I focus directly on solving the i(SM+GR) equations of motion to extract the global structure of the complete vacuum geometry and to interpret its novel physical features. In particular, the flow of information inside black holes is reorganized in ways not previously imagined, as illustrated in the complete causal Penrose diagram of Fig. 11. This has far-reaching implications for the black hole information puzzle and provides new insights into the behavior of the Higgs field, which further enriches the discussion.

The action of the conformally improved i(SM+GR) is [3]

$$S_{i(\text{SM}+\text{GR})} = \int d^4x \sqrt{-g} \left( \begin{array}{c} L_{SM}\left(\psi, A_\mu, g_{\mu\nu}, H\right) + \frac{1}{12}\left(\phi^2 - 2H^\dagger H\right) R\left(g\right) \\ + \frac{1}{2} g^{\mu\nu} \left(\partial_\mu \phi \partial_\nu \phi - 2\partial_\mu H^\dagger \partial_\nu H\right) - V\left(\phi, H\right) \end{array} \right). \quad (1)$$

Here, $\phi, H$ are conformally coupled scalar fields, $g_{\mu\nu}$ is the spacetime metric, and $\psi, A_\mu$ represent the fermions and gauge bosons of the Standard Model. The term $L_{SM}$ includes all familiar fields—quarks, leptons, gauge bosons, dark matter candidates, and right-handed neutrinos—as well as their interactions within SM+GR that are not written explicitly. It also incorporates the remaining parts of the SU(2)xU(1) covariant derivatives $D_\mu H$ in the Higgs doublet kinetic energy term, namely the difference: $\frac{1}{2}g^{\mu\nu}\left(-2D_\mu H^\dagger D_\nu H + 2\partial_\mu H^\dagger \partial_\nu H\right)$. The term $\int L_{SM}$, and the remainder of the full action $S_{i(SM+GR)}$, are separately invariant under local scale (Weyl) transformations with parameter $\Omega(x)$:

$$\phi(x) \to (\Omega(x))^{-1} \phi(x), \ H(x) \to (\Omega(x))^{-1} H(x), \ A_\mu(x) \to (\Omega(x))^0 \ A_\mu(x),$$
$$\psi(x) \to (\Omega(x))^{-3/2} \psi(x), \ g_{\mu\nu}(x) \to (\Omega(x))^2 g_{\mu\nu}(x), \tag{2}$$

This symmetry requires the simultaneous presence of both SM and GR. The Higgs field provides the essential link between the two: it must couple both to the SM and to GR. Thus, *incorporating GR into the SM is not optional but a necessity*, mandated by this special form of local scale symmetry. The two sectors cannot be separated unless the symmetry is broken.

The SU(2)xU(1) singlet $\phi(x)$ is the only field in i(SM+GR) absent in the familiar (SM+GR)[2]. After accounting for the extra local scale gauge symmetry (e.g., by gauge fixing $\phi$ or another degree of freedom), the number of physical (scale-invariant) degrees of freedom in i(SM+GR) equals that of SM+GR. It is therefore natural to view the singlet $\phi$ together with the doublet $H$ as components of a scale-invariant Higgs sector in the ratio $H/\phi$. Accordingly, $\phi$ should be regarded as a gauge extension of the physical Higgs field, and any scale-invariant physical consequences involving $H$ or $\phi$ should be interpreted as arising from the modified Higgs sector comprising both fields.

The conventional Einstein–Hilbert term $(16\pi G_N)^{-1} R(g)$ is excluded from the action (1) because the dimensionful Newton constant $G_N$ violates local scale invariance (2). Instead, a symmetry-preserving *dynamical* gravitational strength $G(x)$, determined by the scalars $\phi(x)$ and $H(x)$, arises in ((1)) as

$$(8\pi G(x))^{-1} \equiv \frac{1}{6} \phi^2(x) \left(1 - h^2(x)\right), \ \text{with} \ h^2 \equiv \frac{2H^\dagger H}{\phi^2}. \tag{3}$$

The factor $(1 - h^2(x))$, central to the following discussion, is invariant under both local scale (Weyl) and SU(2)xU(1) transformations. Therefore, the physical properties of $G(x)$ in various domains are physical and occur in every gauge choice of local scale [3].

---

[2] The local scale symmetry (2) requires the introduction of the scalar field $\phi(x)$, an SU(2)xU(1) singlet, in addition to the Higgs doublet $H$. Both $\phi$ and $H$ must couple conformally to $R(g)$ with a fixed coefficient $1/12$, and with opposite relative signs, as shown in (1) and (3). If $\phi$ were absent—or had the same sign as $H$—the dynamical gravitational strength $G(x)$ in (3) would be negative throughout spacetime $\{x^\mu\}$, contradicting the observed positive Newton constant $G_N$ in our region of the universe. Thus, both the presence of $\phi$ and its relative minus sign are indispensable for simultaneously preserving local scale symmetry and ensuring regions of spacetime where $G(x)$ is positive.

As a side remark, it is worth noting that supergravity (SUGRA) also predics a dynamical, potentially sign changing, gravitational strength $G(x)$ similar to (3), that suggests gravity and antigravity domains. The curvature term in SUGRA is given by $\left((16\pi G_N)^{-1} - \frac{1}{6}K(\varphi, \bar{\varphi})\right) R(g)$ where $K(\varphi, \bar{\varphi})$ is the Kähler potential. Hence, SUGRA exhibits a sign-changing gravitational strength $G(x)$ similar to Eq.(3), taken in the c-gauge $\phi(x) \to \phi_0$. Previous SUGRA literature [6] fixated solely on the positive $G(x)$ spacetime patch, and constrained it to remain positive through a Weyl transformation to the Einstein frame, inadvertently resulting in geodesic incompleteness. SUGRA was elevated to a Weyl-symmetric version that incorporates a complex superfield version of $\phi(x)$ [3]. Hence, SUGRA is geodesically complete, akin to i(SM+GR), provided the presence of the antigravity regions that complete the spacetime are acknowledged. Unfortunately, the supergravity literature chose to ignore the negative sign and assumed that $G(x)$ is positive. Under that assumption, the interesting information about antigravity domains predicted by supergravity remained burried under the rug. However, this aspect SUGRA can now be explored by analogy to the concepts discussed in this paper.

## A. i(SM+GR) far away from singularities

It is crucial to recognize that low-energy physics, as probed in particle accelerators and cosmological observations in regions of weak gravity, corresponds to situations in which $\phi(x)$ is of order $10^{19}$ GeV, while the Higgs field takes its familiar value of about 246 GeV. Consequently, the gauge-invariant dimensionless field $h(x)$ in Eq.(3) is extremely small, of order $h(x) \simeq 10^{-17}$, in such circumstances. Meanwhile, for the observed physical phenomena, the dimensionful dynamical $G(x)$ in Eq.(3) is approximately equal to the standard Newton con-

---

[3] At first sight, the field $\phi$ seems to exhibit ghost-like behavior because of the wrong sign in its kinetic term in (1). However, unitarity is preserved because local scale symmetry compensates for the ghost in any gauge. In particular, one may choose the local scale gauge in which $\phi(x) = \phi_0 \sim 10^{19}$ GeV is constant everywhere in our spacetime patch (though in other patches, or at singularities, it may take different constant values, including zero, see Eq.(17) ). This Weyl gauge is called the *c-gauge* (c for constant) [3]. In this gauge, all fields are labeled with the subscript or superscript $c$—$\phi_c(x), H_c(x), g^c_{\mu\nu}(x)$, etc.—to distinguish them from those in other gauges. In the c-gauge, the action (1) reduces to that of the SM+GR, with $\phi_c(x) = \phi_0$ non-dynamical (hence no ghost), while $H_c, g^c_{\mu\nu}, \psi_c$, etc., remain dynamical.

stant $G_N$ almost everywhere (away from singularities) within our positive-gravity spacetime patch:

$$\text{Weak gravity}: \quad (8\pi G(x))^{-1} \simeq \frac{1}{6}\phi_0^2 = (8\pi G_N)^{-1}, \quad \text{and } h^2(x) \sim 10^{-33}. \tag{4}$$

This behavior is effectively captured in the c-gauge, $\phi(x) = \phi_0$, as described in footnote 3, provided $x^\mu$ is in our gravity patch and far from gravitational singularities. Therefore, in such spacetime regions—such as our own—the difference between i(SM+GR) and the traditional SM+GR becomes insignificant, since the dynamical term $(16\pi G(x))^{-1} R(g)$ effectively reduces to the standard Einstein–Hilbert form $(16\pi G_N)^{-1} R(g)$ with a constant $G_N$.

Moreover, in regions of negligible gravity, i(GR+SM) must be consistent with the purely quartic SM potential $V_4(\phi, H)$ of the renormalizable SM that is also consistent with the local scale symmetry

$$\text{Negligible gravity region}: V(\phi, H) \to V_4(\phi, H) = \frac{\lambda}{4}\left(2H^\dagger H - \alpha^2 \phi^2\right)^2 + \frac{\lambda'}{4}\phi^4. \tag{5}$$

In this setting i(SM+GR), in the c-gauge[3] $\phi(x) \to \phi_0$, produces all the dimensionful parameters in the conventional SM+GR from the single dimensionful source $\phi_0$. These include the Newton constant $G_N$, the Higgs vacuum expectation value (VEV) $\langle 2H^\dagger H \rangle = v_H^2$, the Higgs mass $m_H$, and the cosmological constant $\Lambda$:

$$\begin{array}{l}(8\pi G_N)^{-1} = \frac{\phi_0^2}{6}, \quad (8\pi G_N)^{-1}\Lambda = \frac{\lambda'}{4}\phi_0^4, \\ v_H = \alpha\phi_0 \approx 246 \text{ GeV}, \quad m_H = \sqrt{2\lambda v_H^2} \approx 125 \text{ GeV}.\end{array} \tag{6}$$

As in the SM, all quark, lepton, $W^\pm$&$Z$ gauge boson masses are proportional to $v_H = \alpha\phi_0$, making $\phi_0$ the unique source of all dimensionful parameters. This constitutes a remarkable unification of dimensionful parameters within i(SM+GR).

To reproduce the observed values of $G_N, \Lambda, v_H$ and $m_H$, the relevant dimensionless parameters $\alpha, \lambda, \lambda'$, are fixed as follows:

$$\phi_0 \approx 0.596 \times 10^{19} \text{ GeV}, \quad \alpha \approx 4.13 \times 10^{-17}, \quad \lambda' \approx 8.06 \times 10^{-122}, \quad \lambda \approx 0.129. \tag{7}$$

The tiny magnitudes of $\alpha$ and $\lambda'/\lambda$ highlight the well-known hierarchy puzzle, which remains unresolved in both the traditional SM+GR and in i(SM+GR).

It is important to emphasize that, in the c-gauge[3], at low energies the contribution of $2H_c^\dagger H_c(x)$ in the curvature expressions Eqs.(1, 3) is utterly negligible compared to $\phi_0^2$— suppressed by the staggering factor $\alpha^2 \approx 10^{-33}$. This suppression arises because, in the low-energy regime, the Higgs field $H_c(x)$ typically has magnitude around 246 GeV, consisting of its dominant vacuum expectation value $v_H$ plus small fluctuations $\varphi_H(x)$, interpreted as the Higgs particle in the unitary gauge of SU(2)xU(1): $H_c^\dagger(x) = \left(0,\ (v_H + \varphi_H(x))/\sqrt{2}\right)$. Thus, one may approximate $\left(\phi_c^2(x) - 2H_c^\dagger H_c(x)\right) \approx \phi_0^2$ at low energies. With this simplification, all remaining terms in i(SM+GR) coincide exactly with those of the conventional SM+GR. Therefore, when this minuscule $10^{-33}$ correction is ignored, i(SM+GR) reproduces the low-energy phenomenology of SM+GR with complete accuracy.

In conclusion, i(SM+GR) proves to be just as successful as the conventional SM+GR in describing all known low-energy physics within our familiar spacetime domain, while retaining the deeper consistency required to extend into a completion of spacetime beyond singularities.

### B. i(SM+GR) close to and beyond singularities

The distinction between i(SM+GR) and the conventional SM+GR framework arises from the local scale symmetry displayed in Eq.(2). The sign of the dynamical gravitational strength $G(x)$ as defined in Eq.(3), is governed by the gauge-invariant term $(1 - h^2(x))$ across different spacetime patches:

$$\text{sign}\left[G(x)\right] = \text{sign}\left[1 - h^2(x)\right] = \pm 1. \tag{8}$$

Thus, in all Weyl gauges, the magnitude of the scale-invariant Higgs field $h(x)$ determines whether gravity manifests as attractive or repulsive within a given spacetime region:

$$\begin{aligned} \text{when } |h(x)| &< 1,\ \text{sign}\left[G(x)\right] > 0,\ \text{gravity patch,} \\ \text{when } |h(x)| &> 1,\ \text{sign}\left[G(x)\right] < 0,\ \text{antigravity patch.} \end{aligned} \tag{9}$$

The novel physics of i(SM+GR) becomes prominent in regions where $h^2 \sim 1$, corresponding to intense gravitational fields where $G(x)$ is large. This behavior is particularly pronounced near gravitational singularities, as discussed following Eq.(13). Geodesically complete spacetime regions $\{x^\mu\}$ must include domains where $h^2(x) = 1$ and $h^2(x) > 1$, alongside the familiar low-energy domains where $h^2(x) < 1$. In contrast, traditional SM+GR is limited to only the low-energy domain $h^2(x) < 1$, rendering it incomplete as it cannot account for the physical phenomena in the antigravity domains or the physics at singularities.

By encompassing all regimes, i(SM+GR) provides explicitly the dynamics of geometry and matter across the entirety of geodesically complete spacetime.

Another key departure from SM+GR lies in the potential energy $V(\phi, H)$. Local scale invariance mandates that $V(\phi, H)$ be homogeneous of degree 4, satisfying $V(\Omega^{-1}\phi, \Omega^{-1}H) = \Omega^{-4}V(\phi, H)$. This allows the potential to be expressed as:

$$V(\phi, H) = \phi^4 v(h), \tag{10}$$

where $v(h)$ is an arbitrary function of the gauge-invariant $h(x)$. In weak-gravity regions, $V(\phi, H)$ must approximate the quadratic $V_4(\phi, H)$ with minor corrections, aligning with the Standard Model in the absence of gravity Eq.(5). Thus, in negligible-gravity regimes:

$$\begin{aligned} V(\phi, H) \to V_4(\phi, H) &= \tfrac{\lambda}{4}\left(2H^\dagger H - \alpha^2 \phi^2\right)^2 + \tfrac{\lambda'}{4}\phi^4 + \cdots, \\ v(h) \to v_4(h) &= \tfrac{\lambda}{4}\left(h^2 - \alpha^2\right)^2 + \tfrac{\lambda'}{4} + \cdots \end{aligned} \tag{11}$$

These corrections "$\cdots$" are expected to be minimal, reflecting gravitational effects on Standard Model physics at low energies[4].

In strong-gravity regions, such as near singularities where $h^2(x)$ is of order 1, the function $v(h)$ is not tightly constrained by i(SM+GR) beyond the requirement of a smooth transition to the weak-gravity limit as in Eq.(11). Consequently, significant deviations from $v_4(h)$ may occur in these domains, where the scale-invariant Higgs field $h(x)$ deviates markedly from its low-energy value $h_0 \sim \alpha = 3.31 \times 10^{-17}$.

The detailed structure of $v(h)$ is not central to this paper, which focuses on the asymptotic behavior of $h(x)$ far from singularities. However the nature of $v(h)$ will be critical in a forthcoming study [8]. That work will examine the dynamics of the Weyl-invariant Higgs field $h(x)$ as a function of spacetime coordinates $x^\mu$ near and beyond black hole singularities.

The field $h(x)$ is the sole dimensionless, locally SU(2)xU(1) and scale-invariant Higgs field that is physically observable, in contrast to the gauge-dependent five real scalar fields within the singlet $\phi(x)$ and complex doublet $H(x)$. Both $\phi(x)$ and $H(x)$ are integral to the physical Higgs field $h(x)$. The novel physics described here stems exclusively from this gauge-invariant $h(x)$, making it applicable in any gauge.

---

[4] Given that $h^2(x) \sim 10^{-33}$ is extremely small in low-energy physics, a candidate function $v(h)$ with the desired properties could be envisaged by replacing the constants $\lambda, \alpha, \lambda'$ with functions $\lambda(h), \alpha(h), \lambda'(h)$ that converge to their measured constant values when $h$ is small. Such a formulation was explored in [7].

## III. EQUATIONS OF MOTION

Proceeding further, in this paper, the doublet Higgs $H$ is represented in the standard SU(2)×U(1) unitary gauge-fixed form $H^\dagger(x) = (0, s(x)/\sqrt{2})$, simplifying to $2H^\dagger H \equiv s^2$ and $\partial_\mu H^\dagger \partial_\nu H = \frac{1}{2}\partial_\mu s \partial_\nu s$. Additionally, $L_{\text{SM}}$ is disregarded for the remainder of this paper, as its contribution to the total action is locally scale invariant independently, and it is assumed not to influence the subsequent analysis.

Henceforth, the focus is on the locally scale-invariant interaction of the modified Higgs sector $(\phi, s)$ with the gravitational field $g_{\mu\nu}$. As demonstrated below, these collectively govern the geodesically complete and causal geometry in the interior and exterior of black holes (and similarly before and after the Big Bang). This yields profound implications for the behavior of all matter (including that in $L_{\text{SM}}$) at singularities and generates surprising predictions previously unanticipated. Some global predictions are presented here, while local predictions at singularities will be addressed in a separate paper [8].

The equations of motion derived from $S_{\text{i(SM+GR)}}$ after omitting $L_{\text{SM}}$ are

$$\begin{aligned}
R_{\mu\nu}(g) - \tfrac{1}{2}g_{\mu\nu} R(g) &= (8\pi G(x)) T_{\mu\nu}(\phi, s, g_{\mu\nu}), \\
\tfrac{1}{\sqrt{-g}}\partial_\mu \left(\sqrt{-g} g^{\mu\nu} \partial_\nu \phi\right) &= \left(\tfrac{1}{6}\phi R(g) - \partial_\phi V(\phi, s)\right), \\
\tfrac{1}{\sqrt{-g}}\partial_\mu \left(\sqrt{-g} g^{\mu\nu} \partial_\nu s\right) &= \left(\tfrac{1}{6}s R(g) + \partial_s V(\phi, s)\right),
\end{aligned} \quad (12)$$

where the stress tensor $T_{\mu\nu}(\phi, s, g_{\mu\nu})$ is

$$T_{\mu\nu} \equiv \begin{bmatrix} \tfrac{1}{2}\partial_\mu \phi \partial_\nu \phi - \tfrac{1}{2}\partial_\mu s \partial_\nu s - \tfrac{1}{12}\nabla_\mu \partial_\nu (\phi^2 - s^2) \\ -\tfrac{1}{2}g_{\mu\nu}\left(\tfrac{1}{12}\nabla^2 (\phi^2 - s^2) + V(\phi, s)\right) \end{bmatrix}. \quad (13)$$

From Einstein's equation, $(R_{\mu\nu} - (1/2)g_{\mu\nu}R) = (8\pi G(x))T_{\mu\nu}$ in Eq. (12), it is clear that at spacetime locations $x^\mu$ where the gauge-invariant $(1 - h^2(x))$ vanishes, $G(x)$ diverges (see (3)), causing the curvatures $R_{\mu\nu}(x)$, $R(x)$, and $g_{\mu\nu}$ to explode. Thus, locations $x^\mu$ satisfying $h^2(x) = 1$ ($|\phi(x)| = |s(x)|$) correspond to gravitational singularities. It will be argued that, despite singularities, the geometry is continuous and geodesically complete across them, as are all matter fields. Combining all patches yields a complete and continuous geometry, enabling investigation of old and new physical phenomena and pursuit of measurable effects of this completeness in various patches (such as information flow in black holes).

## IV. COMPLETE GEOMETRY OF VACUUM STATE

The vacuum state is associated with the asymptotic field configurations far from singularities. The aim here is to find solutions to the equations of motion (12, 13) in such asymptotic regimes. Accordingly, assuming that all SM fields vanish asymptotically, except for constant values for $(\phi, s, R)$, we omit all derivatives of $(\phi, s)$, and use the potential $V_4(\phi, s)$ in asymptotic regions. Then, the constants $(\phi, s)$ are determined by the last two equations in (12), which become

$$\begin{aligned}\phi\left(-\tfrac{R}{6} - \lambda\alpha^2\left(s^2 - \alpha^2\phi^2\right) + \lambda'\phi^2\right) &= 0, \\ s\left(\tfrac{R}{6} + \lambda\left(s^2 - \alpha^2\phi^2\right)\right) &= 0.\end{aligned} \quad (14)$$

Although these $(\phi, s)$ equations follow from (12), they equivalently extremize the effective potential $V^0_{\text{eff}}$, which includes not only $V_4$ but also the curvature term

$$V^0_{eff} \equiv -\frac{1}{12}\left(\phi^2 - s^2\right)R + \frac{\lambda}{4}\left(s^2 - \alpha^2\phi^2\right)^2 + \frac{\lambda'}{4}\phi^4, \quad (15)$$

where $R$ is treated as a fixed constant while varying $\phi$ and $s$. Meanwhile, the asymptotic constant $R$ must be computed via the first equation in (12) from a local metric $g_{\mu\nu}(x)$ and its derivatives.

Notably, in this effective potential, the $R$ term is central to the current discussion but absent in the geodesically incomplete conventional SM+GR. In particular, since $\text{sign}(\phi^2 - s^2) = \pm$ corresponds to gravity/antigravity patches, respectively, there are two types of solutions $(\phi_\pm, s_\pm, R_\pm)$, distinguished by $\text{sign}(\phi^2 - s^2) = \pm$. Thus, the asymptotic constant $R \to R_\pm$ may differ across geometrical patches, as the local metric $g_{\mu\nu}(x)$ is not constant.

In the gravitational realm, the measured value of $(\phi_+^2 - s_+^2)/12$ is $(16\pi G_N)^{-1}$, where $G_N$ is the familiar Newton constant. In the antigravity domain, a similar positive constant $\tilde{G}_N$ may be introduced, its value currently unknown. Accordingly,

$$\frac{1}{12}\left(\phi_+^2 - s_+^2\right) = +(16\pi G_N)^{-1}, \quad \frac{1}{12}\left(\phi_-^2 - s_-^2\right) = -\left(16\pi \tilde{G}_N\right)^{-1}. \quad (16)$$

These are not conditions on $(\phi_\pm, s_\pm)$ but definitions of the positive constants $(G_N, \tilde{G}_N)$ guiding the physical interpretation of the solutions $(\phi_\pm, s_\pm, R_\pm)$ below.

### A. Asymptotic scalar fields

The three equations (14) and (16), based solely on the asymptotic $V_4$ (ignoring corrections in $v(h)$ in Eq. (11)), uniquely determine the three dimensionful asymptotic constants $\phi_\pm$,

$s_\pm$, and $R_\pm$ in terms of the gravitational constants $(G_N, \tilde{G}_N)$ in each patch as follows

| gravity patch | antigravity patch |
|---|---|
| $\phi_+^2 = \frac{12}{16\pi G_N} \frac{1-\alpha^2}{(1-\alpha^2)^2 + \lambda'/\lambda} \simeq \frac{6}{8\pi G_N}$, | $\phi_-^2 = 0$ (or $\approx 0$), |
| $s_+^2 = \frac{12}{16\pi G_N} \frac{\alpha^2(1-\alpha^2) - \lambda'/\lambda}{(1-\alpha^2)^2 + \lambda'/\lambda} \simeq \frac{6\alpha^2}{8\pi G_N}$, | $s_-^2 = \frac{6}{8\pi \tilde{G}_N}$, |
| $h_+^2 = \frac{\alpha^2(1-\alpha^2) - \lambda'/\lambda}{1-\alpha^2} \simeq \alpha^2$ | $h_-^2 = \infty$ (or $\approx \infty$) |
| $R_+ = \frac{72}{16\pi G_N} \frac{\lambda'}{(1-\alpha^2)^2 + \lambda'/\lambda} \simeq \frac{+36\lambda'}{8\pi G_N}$, | $R_- = -\frac{36\lambda}{8\pi \tilde{G}_N}$, |
| $V_4(\phi_+, s_+) = \frac{R_+/4}{8\pi G_N} \simeq \frac{9\lambda'}{(8\pi G_N)^2}$, | $V_4(\phi_-, s_-) = \frac{-R_-/4}{8\pi \tilde{G}_N} = \frac{9\lambda}{(8\pi \tilde{G}_N)^2}.$ |

(17)

The approximate expressions in the gravity patch arise from the minute dimensionless constants $\alpha^2 \sim 10^{-33}$, $\lambda'/\lambda \sim 10^{-121}$, given in Eq. (7). In the gravity region, a second solution with $\tilde{s}_+ = 0$ and $\tilde{\phi}_+ \neq 0$ exists but is rejected, as it is not a minimum and does not align with the phenomenologically established SU(2)×U(1) spontaneously broken phase of the Higgs field at low energies. On the antigravity side, $\phi_- = 0$ is the only solution, as the term multiplying $\phi_-$ in (14) becomes positive definite after accounting for the other two equations. Conceivably, $\phi_-$ may have a small nonzero value due to corrections in $V(\phi, s)$ noted in (11). The naive $\phi = s = 0$ solution of (14) is rejected as an asymptotic solution, as it contradicts the existence of the asymptotic weak gravity coupling $G_N \neq 0$ and $\tilde{G}_N \neq 0$.

The only unknown constant in these solutions is the gravitational strength $\tilde{G}_N$ in the antigravity vacuum, while all other parameters are fixed. In particular, the dimensionful Newton's constant $G_N$ and the dimensionless constants $\lambda \approx 0.136$, $\alpha^2 \approx 10^{-33}$, $\lambda'/\lambda \approx 10^{-121}$, are already determined by phenomenological measurements in the gravity sector, as outlined in Eqs. (6, 7). Notably, the curvature $R_-$ is predicted to be inversely proportional to the strength of antigravity, with the proportionality constant

$$R_- = -\frac{36\lambda}{8\pi \tilde{G}_N} \approx -\frac{0.195}{\tilde{G}_N}. \tag{18}$$

Thus, the geometry on the antigravity side is asymptotically anti-de Sitter (AdS).

For later convenience, in the antigravity domain, introduce a dimensionless parameter $0 < \beta < 1$ to characterize simultaneously the antigravity strength and the AdS curvature:

$$8\pi \tilde{G}_N \equiv 3\lambda r_0^2 \frac{\beta^3}{1-\beta}, \quad R_- = -\frac{12}{r_0^2} \frac{1-\beta}{\beta^3}. \tag{19}$$

This dimensionless $\beta$ is the only undetermined parameter in the asymptotic solution displayed in (17).

This establishes that the environment observers like us inhabit corresponds to asymptotic regions of the gravity patch, which is a de Sitter spacetime with tiny positive curvature $R_+$ due to the minute dimensionless constant $\lambda'$. This region is filled with a universal Higgs expectation value $s_+$ of order 246 GeV and is governed by a universal gravitational Newton constant $G_N$. All quark, lepton, and gauge boson masses are proportional to the constant $s_+$. This mirrors the vacuum state setup in traditional SM+GR in asymptotic regions away from singularities, explaining why i(SM+GR) is as effective as traditional SM+GR in accurately describing observed phenomena at low energies and weak gravity[5].

Beyond SMGR, the improved theory i(SM+GR) predicts an antigravity domain with asymptotic AdS geometry. The asymptotic antigravity strength $\tilde{G}_N$ is inversely correlated with the magnitude of the asymptotic AdS curvature $R_-$. Remarkably, the Higgs quartic coupling constant $\lambda$, already known in the gravity region, appears in this correlation in the antigravity region, as seen in (18,19).

### B. Geodesically complete black hole geometry

It remains to determine the continuous geometry $g_{\mu\nu}$ satisfying the first equation in (12) consistententely with the constant curvatures $R_\pm$. Begin with the stress tensor $T_{\mu\nu}$ (13), and evaluate it for the solutions in (17):

$$T_{\mu\nu}^{\pm} = -g_{\mu\nu}^{\pm} \frac{12 V_4(\phi_\pm, s_\pm)}{2(\phi_\pm^2 - s_\pm^2)} = -g_{\mu\nu}^{\pm}\left(8\pi G_N^{\pm}\right) V_4(\phi_\pm, s_\pm) \equiv -g_{\mu\nu}^{\pm}(x)\Lambda_\pm, \tag{20}$$

where $g_{\mu\nu}^{\pm}(x)$ will be computed below. From these, the cosmological constants $\Lambda_\pm$ are identified, after defining the gravity strength $G_N^{\pm}$ in gravity/antigravity sectors as

$$\Lambda_\pm = \left(8\pi G_N^{\pm}\right) V_4(\phi_\pm, s_\pm), \text{ with } G_N^+ \equiv G_N \text{ and } G_N^- \equiv -\tilde{G}_N. \tag{21}$$

Noting $V_4(\phi, s)$ in (5) is strictly positive, the signs of $\Lambda_\pm$ in gravity/antigravity patches are

$$\text{sign}(\Lambda_\pm) = \text{sign}\left(\phi_\pm^2 - s_\pm^2\right) = \text{sign}\left(G_N^{\pm}\right) = \text{sign}(R_\pm) = \text{sign}\left(1 - h_\pm^2\right) = \pm 1. \tag{22}$$

---

[5] By contrast, SM+GR versus i(SM+GR) are very different in regions of strong gravity. Near gravitational singularities, the conformally coupled Higgs sacalars, $\phi(x^\mu)$ and $s(x^\mu)$, exhibit rather unexpected non-perturbative wild behavior as functions of spacetime [8]. Meanwhile, the traditional SM+GR leads one to believe that even at gravitational singularities the Higgs field remains dominated by the same universal constant value that fills the entire universe

This sign, associated with gravity/antigravity patches, is Weyl gauge invariant, since it depends only on the locally scale-invariant Higgs field $h(x) = s(x)/\phi(x)$.

Computing the trace of the Einstein equation, $\left(R^{\pm}_{\mu\nu} - g^{\pm}_{\mu\nu}R^{\pm}/2\right) = -g^{\pm}_{\mu\nu}\Lambda_{\pm}$, yields $\Lambda_{\pm} = R_{\pm}/4$. Thus, per Eq. (17), $\Lambda_{\pm}$ for these solutions is

$$\begin{aligned}
\text{gravity:} \quad & \Lambda_{+} = \tfrac{R_{+}}{4} = \tfrac{9}{8\pi G_N} \tfrac{\lambda'}{(1-\alpha^2)^2 + \lambda'/\lambda} \simeq \tfrac{9\lambda'}{8\pi G_N}, \\
\text{antigravity:} \quad & \Lambda_{-} = \tfrac{R_{-}}{4} = -\tfrac{9\lambda}{8\pi \tilde{G}_N} = -\tfrac{3}{r_0^2}\tfrac{1-\beta}{\beta^3}, \\
\text{ratio:} \quad & \tfrac{\Lambda_{-}}{\Lambda_{+}} = \tfrac{R_{-}}{R_{+}} = \left[\tfrac{\lambda}{\lambda'}(1-\alpha^2)^2 + 1\right]\tfrac{G_N}{\tilde{G}_N} \simeq \tfrac{\lambda}{\lambda'}\tfrac{G_N}{\tilde{G}_N}.
\end{aligned} \quad (23)$$

The first factor in the ratio is huge: $(\lambda/\lambda') \sim 5 \times 10^{120}$. The second factor $(G_N/\tilde{G}_N)$ is unknown.

The antigravity patch predicted by i(SM+GR) does not exist in geodesically incomplete traditional SM+GR. This additional spacetime patch is anti-de Sitter with negative constant curvature $R_{-}$ proportional to $(-\lambda/\tilde{G}_N)$. Unlike the tiny positive $R_{+}$ phenomenologically determined in the familiar gravity patch, the magnitude of negative $R_{-}$ in the antigravity patch is unknown, as the strength $\tilde{G}_N$ (or equivalently Higgs vacuum $s_{-}$, or the $\beta$ parameter) in the antigravity domain is undetermined phenomenologically.

Although $s_{-}$'s overall scale is unknown, *mass ratios in the antigravity region* match identically those in gravity, as masses of quarks, leptons, and $W^{\pm}$, $Z^0$ bosons in both domains are the product of respective mass scales $s_{\pm}$ and the same dimensionless gauge or Yukawa couplings in $L_{SM}$ for traditional SM. Since $\lambda$ is known, anti-de Sitter curvature $R_{-}$ is determined once the Higgs vacuum $s_{-}$ is fixed, or once the mass of any particle in the antigravity sector is determined. Thus, measuring one dimensionful parameter in the antigravity domain for any degree of freedom suffices to fix the dimensionless parameter $\beta$ and determine all dimensionful observables there.

Having determined $T^{\pm}_{\mu\nu}$, now solve for the local metric $g^{\pm}_{\mu\nu}(x)$ satisfying the first equation in (12). Solutions $g^{\pm}_{\mu\nu}(x)$ for constant curvature $R_{\pm}$ in each patch are known, but here gravity and antigravity patches must be joined together to ensure a complete geometry that is continuous at $r = 0$. The solution $g^{\pm}_{\mu\nu}(x)$ meeting this continuity condition is

$$\begin{aligned}
ds^2_{\pm} = -dt^2 A_{\pm}(|\vec{r}|) + \tfrac{1}{A_{\pm}(|\vec{r}|)}(d|\vec{r}|)^2 + \vec{r}^2 d\Omega^2, \\
\text{where } A_{\pm}(|\vec{r}|) = 1 - \tfrac{r_0}{|\vec{r}|}\text{sign}(G^{\pm}_N) - \tfrac{\Lambda_{\pm}}{3}\vec{r}^2.
\end{aligned} \quad (24)$$

Namely, continuity requires that the coefficient of $(-1/|\vec{r}|)$ is $r_0 \text{sign}(G^{\pm}_N)$ as will be further clarified below. Then it is observed that on the gravity side with $\text{sign}(G^{+}_N) = +1$, $A_{+}(|\vec{r}|)$ has a black hole singularity at $\vec{r}=0$, with $r_0 > 0$ interpreted as approximate horizon radius (see Eq. (29)) of a Schwarzschild-de Sitter (SdS) black hole. Parameter $r_0$ relates to black hole mass $M_{bh}$ by

$$r_0 = 2G_N M_{bh} , \qquad (25)$$

where $G_N$ is Newton's constant in familiar gravity spacetime of usual SM+GR. Meanwhile, $A_-(|\vec{r}|)$ is strictly positive due to negative $\text{sign}(\tilde{G}_N^-) = -1$, so there is no value of $|\vec{r}|$ in the antigravity side at which $A_-(|\vec{r}|)$ vanishes. Thus, $A_-(|\vec{r}|)$ describes a negative-mass anti-de Sitter Schwarzschild black hole (adSS) in antigravity patch, with *no horizon*.

Both $A_\pm(|\vec{r}|)$ satisfy the differential equation for any values of the integration constants $\pm r_0$. These integration constants could have been independent of each other for the sake of satisfying the Einstein equation, but they must be correlated in the form $r_0 \text{sign}(G_N^\pm) = \pm r_0$ in order to insure continuity of the spacetime at the singularity as clarified in more detail following Eqs.(27-28).

Gravity and antigravity patches are distinct spacetimes, separated by some boundary that we are about to address. Parametrizing with distinct symbols $x_\pm^\mu = (t_\pm, \vec{r}_\pm)$ distinguishes the coordinates for different domains. The vacuum values of the fields $\phi_\pm, s_\pm$ are their values far from any singularity, $|\vec{r}_\pm| \to \infty$ in different patches. For simplicity, dropping the $\pm$ labels, conventional notation $x^\mu = (t, \vec{r})$ can be maintained for both, as in (24), unless ambiguous.

In $A_\pm(|\vec{r}|)$ in expressions in (24), the symbol $|\vec{r}| = \sqrt{\vec{r} \cdot \vec{r}}$ (i.e., $\sqrt{\vec{r}_\pm \cdot \vec{r}_\pm}$) is the distance from a black hole to the location $\vec{r}_\pm$ in gravity/antigravity sides respectively. Then the $1/|\vec{r}|$ term in $A_\pm(|\vec{r}|)$ can be rewritten by pulling the $\text{sign}(G_N^\pm)$ in front of the square root and adopting a new definition of the spherical coordinate symbol $r$ as follows:

$$\frac{r_0 \text{sign}(G_N^\pm)}{|\vec{r}|} = \frac{r_0}{\text{sign}(G_N^\pm)\sqrt{\vec{r}_\pm \cdot \vec{r}_\pm}} = \frac{r_0}{\text{sign}(1-h^2(x))\sqrt{\vec{r}\cdot\vec{r}}} = \frac{r_0}{r}. \qquad (26)$$

The symbol $r$ in the last expression takes on a new meaning by having both positive and negative values. Thus it spans not only the half real line (deviating from the conventional definition) but the infinite real line:

$$r \equiv |\vec{r}| \; \text{sign}(1 - h^2(x)), \quad -\infty < r < \infty . \qquad (27)$$

This extended $r$ range unifies notation covering regions on opposite sides of the singularity at $r=0$. Then $A_\pm$ and metric $g_{\mu\nu}^\pm$ combine into unified expressions as functions of $r$ in the

extended range

$$ds^2 = -dt^2 A(r) + \frac{1}{A(r)} dr^2 + r^2 d\Omega^2,$$
$$A(r) = \left(1 - \frac{r_0}{r} - \frac{\Lambda(r)}{3} r^2\right), \quad -\infty < r < +\infty, \quad (28)$$
$$\Lambda(r) \equiv \theta(r) \Lambda_+ + \theta(-r) \Lambda_-, \quad r_0 \equiv 2G_N m_{bh}.$$

Note $A(r)$ or $(A(r))^{-1}$ and metric $g_{\mu\nu}$ are singular over the full range $-\infty < r < +\infty$, but are continuous at $r = 0$ (see Figs.1&2). This mathematical notation, that physically corresponds to the presence of the gravity/antigravity patches on opposite sides of the singularity, provides a mathematical setting for a *manifold that unifies the gravity/antigravity patches into a single unified spacetime* for the same blackhole with mass $M_{bh}$.

This point was first recognized in [9] for vanishing curvature, and it is generalized here to include nontrivial $R_\pm$. Here, the Higgs field $h(x)$ necessarily creates the cosmological constants $\Lambda_\pm$ that shape the asymptotic geometry of the spacetime on both the gravity and antigravity regions. The current paper introduces (27) as a notation change for $r$, but the substance is unchanged when the $\Lambda_\pm \to 0$ geometry is expressed in terms of Kruskal-Szekeres coordinates $(u, v)$ in this paper versus [9].

In the case of $r_0 = 0$ or $M_{bh} = 0$, when there is no black hole singularity, the geometry is still continuous at $r = 0$, but the boundary region $r \sim 0$ that separates them does not have a distinctive geometrical shape. However, when continuous local fields $\phi(x^\mu)$ and $s(x^\mu)$ are taken into consideration in the vicinity of the black hole, there still exists a black hole singularity where gravity transitions into antigravity even in the limit $r_0 \to 0$ (see [8]). In that setting, the $r_0 \to 0$ limit is a continuous solution with a black hole.

From here on, the new black hole solution in (28) will be referred to as the AdSSdS black hole. In the $r > 0$ gravity side it coincides with the well known SdS black hole, in the $r < 0$ antigravity side, as a function of $|\vec{r}|$ (after replacing $r = -|\vec{r}|$) it coincides with the horizonless Anti-deSitter-Schwarzchild (AdSS) black hole with *negative mass*. This AdSSdS is a new solution that is possible only in the unified manifold that includes gravity and antigravity patches, which is predicted by the improved theory i(SM+GR).

### C. Some physical features of the complete AdSSdS geometry

*Horizons from $A_+(r) = 0$:* The equation $A_+(r) = 0$ for $r > 0$ is cubic, with three roots of physical and geometrical significance. Two are positive and correspond to the black hole horizon $r_h$ and the cosmological horizon $r_c$, as shown in Figs. 1&2. The third root, $-(r_h + r_c)$, is negative and irrelevant for $r > 0$. For $r < 0$, solving $A_-(r) = 0$ gives two complex conjugate roots $w, w^*$ and one real positive root $-(w + w^*)$. Thus $A_-(r)$ never vanishes in the $r < 0$ region, as illustrated in Figs. 1&2.

*Plots of $A(r)$ and $A(r)^{-1}$:* The functions $A(r)$ and $A(r)^{-1}$ over the full range $-\infty < r < \infty$, as defined in Eq.(28), are plotted in Figs. 1&2 using unphysical numerical values of $(r_h, r_c, w)$ to highlight qualitative features. The plots show that both $A(r)$ and $A(r)^{-1}$ are continuous at the singularity $r = 0$ and at the horizons $r_h, r_c$. Notably, $A(r)$ is strictly positive in the antigravity region ($r < 0$), while in the gravity region ($r > 0$) its sign alternates.

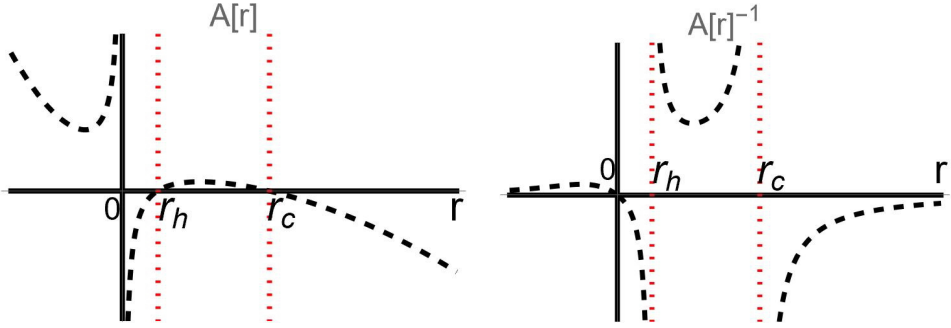

Fig.1- $A(r)$ vanishes at $r_h$ & $r_c$.  Fig.2- $A(r)^{-1}$.vanishes at $r = 0$.

*Formulas for horizons and roots:* The quantities $r_h, r_c, w$ can be expressed in terms of $r_0$ and $\Lambda_\pm$:

$$\begin{aligned} r_h &= \tfrac{2r_0}{\sqrt{\Lambda_+ r_0^2}} \cos\left(\tfrac{1}{3}\cos^{-1}\left(\tfrac{3}{2}\sqrt{\Lambda_+ r_0^2}\right) + \tfrac{\pi}{3}\right) \simeq r_0 \left(1 + \tfrac{\Lambda_+ r_0^2}{3} + \cdots\right), \\ r_c &= \tfrac{2r_0}{\sqrt{\Lambda_+ r_0^2}} \cos\left(\tfrac{1}{3}\cos^{-1}\left(\tfrac{3}{2}\sqrt{\Lambda_+ r_0^2}\right) - \tfrac{\pi}{3}\right) \simeq \tfrac{\sqrt{3}r_0}{\sqrt{\Lambda_+ r_0^2}} \left(1 - \tfrac{\sqrt{\Lambda_+ r_0^2}}{2\sqrt{3}} + \cdots\right), \\ w &= \tfrac{-2r_0}{\sqrt{-\Lambda_- r_0^2}} \sinh\left(\tfrac{1}{3}\sinh^{-1}\left(\tfrac{3}{2}\sqrt{-\Lambda_- r_0^2}\right) + i\tfrac{\pi}{3}\right). \end{aligned} \qquad (29)$$

The approximations for $r_h, r_c$ are useful because the dimensionless factor $\Lambda_+ r_0^2$ is extremely small, due to the minute $\lambda' \sim 10^{-122}$:

$$\Lambda_+ r_0^2 \simeq \frac{3r_0^2}{r_c^2} \simeq \frac{9\lambda' r_0^2}{8\pi G_N} \approx 5.4 \times 10^{-35} \left(\frac{r_0}{r_\odot}\right)^2. \qquad (30)$$

Here $r_\odot$ is the solar radius, about $0.7 \times 10^9$ m. Thus the black hole horizon radius satisfies $r_h \approx r_0$, while the cosmological horizon is enormous—of order the visible universe:

$$r_c \simeq \frac{\sqrt{3}}{\sqrt{\Lambda_+}} = \sqrt{\frac{8\pi G_N}{3\lambda'}} \approx 1.65 \times 10^{26} \text{ m} \quad (17.41 \text{ billion light years}) \tag{31}$$

It follows that the ratio $r_c/r_h$ is extremely large:

$$\frac{r_c}{r_h} \simeq \frac{\sqrt{3}}{\sqrt{\Lambda_+ r_0^2}} = 2.36 \times 10^{17} \times \frac{r_\odot}{r_0}. \tag{32}$$

The de Sitter curvature can also be expressed in terms of $r_c$ since $R_+ = 4\Lambda_+ \simeq 12/r_c^2$, which is tiny since $r_c$ is huge.

*Visible universe region:* We observe physical phenomena only in the bounded visible region, outside black hole horizons $(r > r_h)$ but inside the cosmological horizon $(r < r_c)$:

$$\text{visible universe:} \quad r_h < r < r_c. \tag{33}$$

From Eq.(31), the estimated diameter of the visible universe is $2r_c \approx 34.82$ billion light years. An independent estimate [10], based on the age of the universe (13.8 billion years), and taking into account an initial inflationary period, yields $\sim 93$ billion light years. While both results are of the same order of magnitude, the ratio $93/(2r_c) = 2.67$ indicates a discrepancy. It would be of interest to resolve this tension since the number 93 relies indirectly on an early universe inflation while the number $2r_c = 34.82$ relies on the direct measurement of the cosmological constant for current universe.

Thus, our expanding universe is much larger than the visible portion. Beyond $r_c$, as well as in the interior regions $-\infty < r < r_h$ of all black holes, lies spacetime we cannot currently observe directly. In a geodesically complete universe extending to the pre–Big Bang era as well, further regions with additional black holes are expected. In fact, i(SM+GR) suggests a cyclic universe scenario driven by Higgs field dynamics: an infinite sequence of big crunches and big bangs, separated by brief antigravity phases [7]. The improved theory i(SM+GR) achieves geodesic completeness by encompassing all such regions.

*Relations among parameters*: To make the structure of $A(r)$ and $A(r)^{-1}$ explicit, we rewrite them in forms where $(r_h, r_c, w)$ are primary parameters, while $(r_0, \Lambda_\pm)$ are derived:

$$A(r) = \begin{cases} \theta(r)\left[-\frac{\Lambda_+}{3r}(r-r_h)(r-r_c)(r+r_h+r_c)\right] \\ +\theta(-r)\left[-\frac{\Lambda_-}{3r}(r-w)(r-w^*)(r+w+w^*)\right] \end{cases},$$
$$(A(r))^{-1} = \begin{cases} \theta(r)\left[\frac{\rho_h}{r-r_h}+\frac{\rho_c}{r-r_c}-\frac{\rho_h+\rho_c}{r+r_h+r_c}\right] \\ +\theta(-r)\left[\frac{\rho_w}{r-w}+\frac{\rho_w^*}{r-w^*}-\frac{\rho_w+\rho_w^*}{r+w+w^*}\right] \end{cases}. \tag{34}$$

These forms are useful for constructing Kruskal–Szekeres coordinates (section V).

Expanding $A(r)$ in Eq.(34) and comparing with Eq.(28) yields $r_0, \Lambda_\pm$ in terms of $(r_h, r_c, w)$, and allows evaluation of $(\rho_h, \rho_c, \rho_w)$:

$$\begin{aligned} \frac{\Lambda_+}{3} &= \left(r_c^2 + r_c r_h + r_h^2\right)^{-1}, \quad \frac{\Lambda_-}{3} = \left(w^2 + ww^* + (w^*)^2\right)^{-1} < 0, \\ r_0 &= \frac{r_c r_h(r_c+r_h)}{r_c^2+r_c r_h+r_h^2} = \frac{ww^*(w+w^*)}{w^2+ww^*+(w^*)^2} > 0, \text{ hence } (w+w^*) < 0, \\ \rho_h &= \frac{r_h(r_c^2+r_c r_h+r_h^2)}{(r_c-r_h)(r_c+2r_h)} = \frac{r_h}{1-\Lambda_+ r_h^2} \approx r_h\left(1+3\frac{r_h^2}{r_c^2}+\cdots\right), \\ \rho_c &= \frac{-r_c(r_c^2+r_c r_h+r_h^2)}{(r_c-r_h)(2r_c+r_h)} = \frac{r_c}{1-\Lambda_+ r_c^2} \approx -\frac{r_c}{2}\left(1+\frac{3}{2}\frac{r_h}{r_c}+\cdots\right), \\ \rho_w &= w\frac{(w^2+ww^*+(w^*)^2)}{(w^*-w)(w^*+2w)} = \frac{w}{1-\Lambda_- w^2}. \end{aligned} \tag{35}$$

The quantities $(2\rho_h)^{-1}, (-2\rho_c)^{-1}$ have direct physical significance: they are the surface gravities $(\kappa_h, \kappa_c)$ at the black hole and cosmological horizons, respectively [11]:

$$\kappa_{h,c} \equiv \frac{1}{2}(dA/dr)^{-1}_{r=r_h,r_c} = (2|\rho_{h,c}|)^{-1}. \tag{36}$$

*Antigravity region and $\beta$-parametrization:* Next, consider the antigravity region. To satisfy the $r_0$ condition (second line of Eq.(35)), it is convenient to reparametrize $w$ (from Eq.(29)) and $(\rho_w, \Lambda_-)$ (from Eq.(35)) in terms of $r_0$ and the dimensionless parameter $\beta$ introduced in Eq.(19):

$$\begin{aligned} w &= r_0 \frac{\beta}{2}\left(-1 - i\sqrt{\frac{3+\beta}{1-\beta}}\right), \quad 0 \le \beta \le 1, \\ \rho_w &= r_0 \frac{\beta^2}{6-4\beta}\left(-1 + i\sqrt{\frac{3+\beta}{1-\beta}\frac{3-\beta}{3+\beta}}\right). \end{aligned} \tag{37}$$

In contrast to the approximations available for $(r_h, r_c, \Lambda_+ r_0^2)$ and $(\rho_h, \rho_c)$ noted in (29,30,35), no analogous approximations exist for the dimensionless quantities $w(\beta)$ and $\rho_w(\beta)$ and related quantities $\Lambda_-(\beta), R_-(\beta), \tilde{G}_N(\beta)$, all of which depend on a single unknown $\beta$. However, the possible range of numerical values of these quantities can be plotted as the dimensionless $\beta$ (equivalently $\tilde{G}_N$) changes in the range $0 \le \beta \le 1$. In particular: as $\beta \to 0$, antigravity is weak ($\tilde{G}_N \to 0$), and both $w$ and $\rho_w$ vanish, and as $\beta \to 1$, antigravity is strong ($\tilde{G}_N \to \infty$), and $w, \rho_w$ diverge along their respective imaginary directions. The precise limiting behaviors are:

|  | $\beta \approx 0$ | $\beta \approx 1$ |
|---|---|---|
| | $8\pi \tilde{G}_N \approx \beta^3 3\lambda r_0^2 (1 + \cdots)$ | $\tilde{G}_N \approx \frac{1}{(1-\beta)} \frac{3\lambda r_0^2}{8\pi} + \cdots$ |
| | $R_- \approx \frac{-1}{\beta^3} \frac{12}{r_0^2} + \cdots$ | $R_- \approx -(1-\beta) \frac{12}{r_0^2} + \cdots$ |
| | $\Lambda_- r_0^2 \approx -\frac{3}{\beta^3} + \cdots$ | $\Lambda_- r_0^2 \approx 3(1-\beta) + \cdots$ |
| | $\frac{w}{r_0} \approx \beta \left( e^{-i\frac{2\pi}{3}} - i\frac{\beta}{\sqrt{3}} \right) + \cdots$ | $\frac{w}{r_0} \approx -\frac{i}{\sqrt{1-\beta}} - \frac{1}{2} + \cdots$ |
| | $\frac{\rho_w}{r_0} \approx \frac{\beta^2}{3} e^{i\frac{2\pi}{3}} + \cdots$ | $\frac{\rho_w}{r_0} \approx \frac{i}{2\sqrt{1-\beta}} - \frac{1}{2} + \cdots$ |

(38)

These limiting forms are particularly useful for constructing the Kruskal–Szekeres geometry of the antigravity spacetime and for analyzing the flow of information across and beyond the black hole singularity, as discussed in section V.

### D. Complete AdSSdS black hole geodesics bridging gravity-antigravity

Signals that originate in our visible region $r_h < r < r_c$ can propagate both to the interior black hole region $-\infty < r < r_h$ (including negative $r$ beyond the singularity) and to the exterior cosmological region $r_c < r < \infty$. Proper observers, such as infalling particles or observers receding toward cosmological distances, can indeed cross either horizon and reach the inner or outer regions. However, laboratory observers restricted to the visible region never see such passages. Likewise, signals from the inner or outer regions are inaccessible to laboratory observers because of the causal separation enforced by the horizons.

This effect can be understood by relating laboratory spacetime coordinates $x^\mu = (t, r)$ in Eq.(28) to the proper time $\tau$ via the geodesic embedding $x^\mu(\tau) = (t(\tau), r(\tau))$. Computing the geodesics $x^\mu(\tau)$ shows that, at finite proper time $\tau_h$ or $\tau_c$ when the particle reaches a horizon $r(\tau_h) = r_h$ or $r(\tau_c) = r_c$, the laboratory time diverges: $t(\tau_h) = t(\tau_c) = \infty$ (see Eq.(45)). Thus, while particles do cross horizons within finite proper time, laboratory observers can never witness such crossings, regardless of how long they wait.

The geodesics $x^\mu(\tau)$ of a moving particle follow from the on-shell constraint for a massive or massless particle in curved space,

$$g^{\mu\nu}(x(\tau)) p_\mu(\tau) p_\nu(\tau) + m^2 = 0. \tag{39}$$

with $p_\mu(\tau) \equiv g_{\mu\nu}(x(\tau))\dot{x}^\nu(\tau)$ the momentum, and $x^\nu(\tau) = (x^0(\tau), r(\tau))$ the worldline trajectory as a function of proper time (or affine parameter for massless particles). For any metric $g_{\mu\nu}(x)$ independent of $t \equiv x^0$ and the direction of $\vec{r}$ (angles $\theta, \phi$) there exist conserved quantities: namely, the energy $p_0$, conjugate to $t$, and the angular momentum components $(p_\theta, p_\phi)$, conjugate to $(\theta, \phi)$. Since the metric in Eq.(28) is diagonal in spherical coordinates with determinant 1, the conserved quantities simplify to the energy

$$E \equiv -p_0 = -g_{00}\dot{t}(\tau) = A(r(\tau))\dot{t}(\tau), \tag{40}$$

and the conserved orbital angular momentum $\vec{L} = \vec{r} \times \vec{p}$. The remaining dynamical degrees of freedom are the radial coordinate $r(\tau)$ and its conjugate radial momentum $p_r(\tau) = g_{rr}\dot{r} = \dot{r}/A(r)$. Substituting these into the constraint (39) and multiplying through by $A(r)$, one obtains a nonrelativistic Hamiltonian form (see See Eqs.(25.16a and 25.16b) in reference [14]),

$$\left(\left(\frac{dr}{d\tau}\right)^2 + V_{eff}(r(\tau))\right) = E^2, \quad \text{with } V_{eff}(r) \equiv A(r)\left(\frac{\vec{L}^2}{r^2} + m^2\right). \tag{41}$$

This formulation describes a one-dimensional mechanical system: a particle of "energy" $E^2$ moving along the real line $-\infty < r < \infty$ in the potential $V_{\text{eff}}(r)$. The analogy is a ball rolling along a road whose height profile represents $V_{\text{eff}}(r)$ as the gravitational potential. Turning points occur at $r_k$ where $V_{\text{eff}}(r_k) = E^2$. Between turning points, $r(\tau)$ evolves monotonically; at each turning point, the direction reverses. If no turning points exist, the motion continues indefinitely.

Analytically, the solutions follow from integrating Eq.(41) in regions of $r(\tau)$ that satisfy $E^2 > A(r)\left(\frac{\vec{L}^2}{r^2} + m^2\right)$:

$$\begin{aligned}
\frac{dr}{d\tau} &= \pm\sqrt{E^2 - A(r)\left(\frac{\vec{L}^2}{r^2} + m^2\right)}, \\
\Rightarrow (\tau - \tau_k) &= \pm \int_{r_k}^{r(\tau)} dx \left(E^2 - A(x)\left(\frac{\vec{L}^2}{x^2} + m^2\right)\right)^{-1/2}.
\end{aligned} \tag{42}$$

Here $r_k(E, L, m)$ are turning points, and $\tau_k$ are the corresponding proper times. The sign switches at each turning point while imposing continuity of the solution at each $r_k$.

Once $r(\tau)$ is deduced from the above, the laboratory time follows from Eq.(40):

$$t(\tau) = E \int_0^\tau \frac{d\tau'}{A(r(\tau'))}, \tag{43}$$

where the integral is evaluated using the solution $r(\tau)$. This procedure generates all complete geodesics.

All solutions $r(\tau)$ are readily deduced intuitively, without any computation, by observing a plot of the effective potential $V_{eff}(r)$ as a function of $r$ (see Figs.3,4,5), and marking the corresponding turning points $r_k$ that satisfy $V_{eff}(r_k) = E^2$ for some fixed values of $E^2, L^2, m^2$. Then, the position of the ball $r(\tau)$ increases (or decreases) steadily until it hits a turning point $r_k$ if any, at which point it switches direction and moves again steadily until the next turning point, and so on. If there are no real values for the turning points $r_k(E, L, m)$, then the ball at $r(\tau)$ continues to move indefinitely in the same initial direction.

The procedure described above is a standard computation of geodesics in a black hole when $r > 0$. The novelty in this paper is that this approach gives all the geodesics in the full spacetime, $-\infty < r < \infty$, which includes the continuously connected gravity as well as the antigravity patches of a black hole.

The special cases of $(m^2 \neq 0, L^2 = 0)$, $(m^2 = 0, L^2 \neq 0)$ and $(m^2 = 0, L^2 = 0)$ are insightful. The corresponding plots of $V_{eff}(r)$ are given in Figs.3, 4 and 5 respectively, for some numerical values of $E, m, L$, and using the $A(r)$ plotted in Fig.1&2 (where the black hole and cosmological horizons, $r_h$ and $r_c$, are indicated). The plot for $(m^2 \neq 0, L^2 \neq 0)$ has similar features to Figs.3&4, so it will not be described separately. The horizontal dashed lines in Figs.3,4,5 correspond to increasing values of $E^2 > 0$ represented in different colors. The turning points $\{r_k\}$ are the intersections of some dashed line with the solid blue curve (including the vertical axis in Figs.3&4 but not in Fig.5). Accordingly, the turning points change as $E^2$ changes in Figs.3&4, while there are no turning points at all in Fig.5 since $V_{eff}(r) = 0$ in this case (the blue horizontal line). In the following I will first discuss the case of massless particles with vanishing angular momentum illustrated in Fig.5.

After that discussion, I will address the case of massive particles that require additional input that is not incorporated in Figs.4&5.

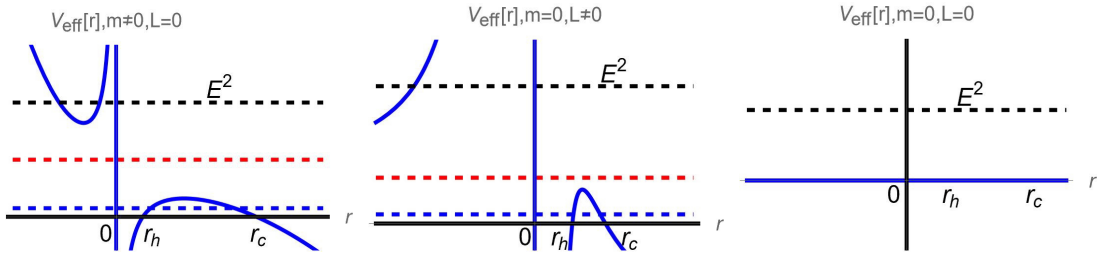

Fig.3 - $m^2 \neq 0$, $L^2 = 0$.    Fig.4 - $m^2 = 0$, $L^2 \neq 0$.    Fig.5 - $m^2 = 0$, $L^2 = 0$.

1. Massless AdSSdS geodesics

For $m = 0$ and $L = 0$, the potential vanishes: $V_{\text{eff}}(r) = 0$ (Fig.5). Hence no turning points exist and Eq.(42) integrates directly:

$$\dot{r}_{m=0} = \mp E, \quad r_{m=0}(\tau) = \mp E\tau, \tag{44}$$

with sign choice determined by the initial direction of the radial velocity $\dot{r}$, while $E$ is positive/negative for particles/antiparticles respectively. For example, $r_{m=0}(\tau) = -E\tau$, with $-\infty < \tau < \infty$, is the complete geodesic of a photon, with $E > 0$ and $\dot{r} = (-E) < 0$, that moves in the direction of decreasing radial position, which means from the gravity region (where $\tau$ is negative) to the antigravity region (where $\tau$ is positive). This geodesic hits the singularity $r = 0$ at $\tau = 0$ (by choice of $\tau$-origin due to $\tau$-translation symmetry). If this photon originates at radial position $r_1 = -E\tau_1$ in the visible region $r_h < r_1 < r_c$ (see Fig.5), it crosses the horizon and reaches the singularity in a finite amount of proper time $|\tau_1| = r_1/E$; it smoothly crosses the singularity and proceeds into the antigravity; it reaches $r \to -\infty$ only as $\tau \to +\infty$.

Similarly, $r_{m=0}(\tau) = E\tau$ is the complete geodesics of photons with positive velocity that can cross the cosmological horizon $r_c$ and proceed to $r \to +\infty$ at infinite proper (or affine) time.

The radial photon geodesics are continuous across both horizons and through the singularity, since $V_{\text{eff}} = 0$. Hence photons, gluons, and gravitons traverse black hole singularities unimpeded, carrying information between gravity and antigravity sectors. The boundaries at $r \to \pm\infty$ are reached only at infinite affine time, implying *geodesic completeness*.

It should be emphasized that the massless radial geodesics $r_{m=0}(\tau) = \mp E\tau$ are valid for all $A(r)$, not just the AdSSdS black hole discussed here. This is because $V_{eff}(r)$ vanishes for any $A(r)$ as long as $m = 0 = L$.

By contrast, the laboratory time $t_{m=0}(\tau)$ does depend on $A(r)$, and it is seen to diverge at finite proper times $\tau_h$ and $\tau_c$ when the photon reaches the horizons at $r_h$ or $r_c$:

$$t_{m=0}(\tau) = E \int_0^\tau d\tau' \left(A\left(r_{m=0}(\tau')\right)\right)^{-1} = \mp \int_0^{r_{m=0}(\tau)} dx \left(A(x)\right)^{-1} = \mp r_*\left(r_{m=0}(\tau)\right), \tag{45}$$

where $x \equiv \mp E\tau'$, and $r_*(r)$ is the tortoise coordinate defined in Eqs.(47,50). The divergence $r_*(r_h) = \infty = r_*(r_c)$ is recorded in the plot in Fig.6. This explains why, laboratory observers located in the visible region $r_h < r < r_c$, can never see the passage of infalling or outgoing light (or other information) through the horizons $r_h$ or $r_c$, no matter how long they wait. This is why, for such observers, the interiors of black holes are invisible, and so is the region of the universe beyond the cosmological horizon.

By contrast, for proper observers (such as the massless particles themselves) time is simply the proper (affine) time $\tau$ while position is $r_{m=0}(\tau)$ (positive and negative) as measured from the center of the black hole in the gravity/antigravity regions. Hence idealized pointlike and massless proper observers (such as photons) do experience the passage through the horizons, as well as through the singularity. The permissible causal direction of such flows of information, especially through the singularity, will be encoded in the causal structure of the Penrose diagram in section V.

However, this is not the full story of geodesic completeness. As will become clearer in section V, the Kruskal Szekeres diagrams in Figs.9&10, or the Penrose diagram in Fig.11, reveal that there exists an extended spacetime that cannot be captured with only the $(t,r)$ parametrization of the gravity+antigravity spacetime. When this *extended gravity+antigravity* spacetime is taken into account, it is seen that there is a bounce at the antigravity boundary $r = -\infty$. Namely, a geodesic that reaches the antigravity boundary at $r = -\infty$ does not end, but gets reflected back into the antigravity region, passing through the singularity to reach into a second gravity region (region $IV$ in the KS digram in Fig.9), and then ending its journey at the asymptotic boundary $r \to \infty$ after passing the cosmological horizon of the second gravity region. This added path in the extended gravity+antigravity spacetime completes the geodesic for a massless particle that falls into a black hole. The details are in section V.

In addition to the solution above for a photon that traverses the singularity, namely $r = -E\tau$, there is a second solution given by $r(\tau) = E|\tau|$. In this case $r$ remains positive for all proper time $-\infty < \tau < \infty$, so it never traverses the singularity. After reaching $r = 0$ in a finite amount of proper time, as above, gets reflected at the center of the black hole at proper time $\tau = 0$ (like a mirror), bouncing back toward increasing $r$, out of the horizon and then entering the second gravity domain (region $IV$ in the KS diagram in Fig.9), and moving to its asymptotic region at $r \to \infty$. So its eventual fate is similar to the other solution. This geodesic is also geodesically complete in i(SM+GR), but is not available in the traditional SM+GR because that incomplete theory gives up describing the physics at proper time $\tau = 0$.

This type of completeness is one of the many mathematical and physical novelties in i(SM+GR) that is absent in the geodesically incomplete (SM+GR).

### 2. Massive AdSSdS geodesics

For $m^2 \neq 0$ or $L^2 \neq 0$, the effective potential introduces barriers at turning points, including at $r = 0$ (Figs.3&4). Classical trajectories bounce at these turning points and cannot classically penetrate into the antigravity sector. However, in quantum theory tunneling occurs under barriers, thus connecting continuously the incoming, transmission and reflection probability amplitudes in the gravity and antigravity regions. Hence, quantum theory inherits and alters somewhat the classical geodesic completeness.

This is not the end of the discussion for massive particles, because the origin of mass in i(SM+GR) is the extended Higgs field $(\phi, s)$. In geodesics computation, such as (41,42), the mass $m$ of quarks, leptons $W^{\pm}, Z^0$ gauge bosons, should be replaced by the $r$-dependent effective *local masses* $m(r) = gs(r)$, where $s(r)$ is the Higgs field, while $g$ is a dimensionless coupling constants (Yukawa or gauge coupling) determined by $L_{SM}$, the traditional Standard Model (SM+GR). This $r$ dependence of $m(r)$ may have a striking effect on traversing the singularity, classically or quantum mechanically, due to a surprizing behavior of the Higgs fields $(\phi, s)$ near the singularity at $r = 0$.

As detailed in [8], beyond their asymptotic values in (17), continuous solutions $\phi(r), s(r), A(r)$ across $-\infty < r < \infty$ show that at the singularity both scalars vanish, while their ratio remains finite

$$\phi(r \to 0) = 0, \ s(r \to 0) = 0, \ \left|\frac{s(r \to 0)}{\phi(r \to 0)}\right| = 1. \qquad (46)$$

Hence the mass vanishes, $m(r \to 0) = 0$, at the singularity. Similar behavior of the Higgs field near cosmological singularities was previously noted in [12][13] (as a function of conformal time instead of as function of $r$). Accordingly, the gauge symmetry SU(2)×U(1) gets re-established precisely at the singularity.

The modified $V_{eff}(r)$ plotted Figs.3&4 would then be different, and may become similar to Fig.5 in the neighborhood of $r = 0$, since $m(0) = 0$ for all the fields of the Standard Model (not only photons gravitons and gluons, but also all quarks, leptons and gauge bosons). It appears that the classical radial geodesics with $L^2 = 0$, for all massless or massive particles of the Standard model (and their quantum probability amplitudes), could perhaps traverse the singularity, just like the massless particle's case illustrated in Eqs.(44,45), when applied to the neighborhood of the singularity. However, this is unclear until further investigation, because in the presence of dynamical $(\phi(r), s(r))$, the profile for the geometry $A(r)$ is also altered, becoming more singular as compared to the standard black hole case in (28).

Therefore, $V_{eff}(r)$ in Eq.(41) needs to be carefully computed before this question about massive geodesics can be settled at the classical level. In any case, at the quantum level, quantum tunelling of probability amplitudes as well as fields into the antigravity domain should be expected.

This effect due a dynamical mass $m(\tau)$ was demonstrated in the context of cosmology analytically, by showing that particle geodesics whose masses are proportional to the Higgs field, do indeed sail through cosmological singularities even in the presence of anisotropy [13]. This notion, that is just emerging for fields near black hole singularities, deserves detailed analysis in future investigations (see [8]).

## V. GEODESICALLY COMPLETE ADSSDS IN $(u, v)$ KS SPACETIME

The flow of information traced by classical geodesics in different gravity and antigravity regions, as discussed previously, can be clarified by examining the broader causal structure of the extended spacetime. This requires analyzing the geodesically complete extension of classical gravity/antigravity geometry in this section. This prepares the grounds for the next section where the Kruskal–Szekeres (KS) and Penrose diagrams reveal the causal structure of the complete spacetime.

In traditional black hole studies, the KS coordinates $(u, v)$ [14] are functions of $(t, |\vec{r}|)$ that unify multiple patches of spacetime which otherwise appear disconnected in the $(t, |\vec{r}|)$ description. These patches are conventionally labeled by $i = (I, II, III, IV)$, as shown in Figs. 9&10. The coordinate maps $(u_i(|\vec{r}|, t), v_i(|\vec{r}|, t))$ differ for each patch, but once the metric $g_{\mu\nu}$ is expressed in terms of $(u, v)$, all regions are continuously and analytically connected through a single analytic function of $(u, v)$, evaluated at different coordinate values of $(u, v)$.

The natural domain of $(u, v)$ is the entire plane $\mathbb{R}^2$, i.e. $-\infty < u < \infty$ and $-\infty < v < \infty$.

However, the traditional black hole spacetime is restricted to the region $uv < 1$, which corresponds to the union of regions $(I, II, III, IV)$ in a KS diagram [15]. This restriction arises because the radial distance from the singularity at $r = 0$, defined as $|\vec{r}| \equiv \sqrt{\vec{r} \cdot \vec{r}}$, must remain positive, $|\vec{r}| > 0$. In KS coordinates this translates to the condition $uv < 1$. The boundary of the traditionally allowed region $uv < 1$ appears as the wavy red hyperbola labeled $r = 0$ or $uv = 1$ in region $II_h$ of the KS diagram in Fig. 9. Consequently, the regions $V$ and $VI$, corresponding to $uv > 1$, are excluded. The mathematical expression for the metric $g_{\mu\nu}(u, v)$, however, is not intrinsically aware of this restriction, so the constraint $uv < 1$ must be imposed by hand. In fact, geodesics in the $(u, v)$ spacetime can reach the singularity at $uv = 1$ in finite proper time $\tau$. Within standard theoretical frameworks, no prescription exists for continuing these geodesics as proper time continues to tick. Thus the spacetime is geodesically incomplete. This incompleteness prevents any consistent description of physics at or beyond the singularity.

One might be tempted to include regions $V$ and $VI$, since they naturally emerge mathematically from the $(u, v)$ formalism. However, such an extension conflicts with the positivity of the distance $|\vec{r}| = \sqrt{\vec{r} \cdot \vec{r}}$, which becomes incompatible with those regions. Hence, in conventional (SM+GR) or its standard extensions in quantum gravity formalisms, such a mathematical analytic continuation, of the $(u, v)$ spacetime beyond $uv = 1$, has no physical meaning.

This is precisely where the new physics of i(SM+GR) provides the missing element, by creating additional spacetime where there is antigravity and the physical phenomena that occur within it. The key lies in the generalized definition, $r = |\vec{r}| \, \text{sign}(1 - h^2)$, introduced in Eq.(27) and used in the geodesically complete metric of Eq.(28). In this framework, both $t$ and $r$ span the full real line, $-\infty < t < \infty$ and $-\infty < r < \infty$. Extending $r$ from the half-line to the full line is tied to the nonperturbative dynamics of the locally scale-invariant Higgs field $h(x)$, which interpolates between the gravity regime $h^2(r > 0) < 1$ and the antigravity regime $h^2(r < 0) > 1$. When translated into KS coordinates through Eqs.47-50), the negative-$r$ region, corresponding to antigravity, maps precisely to regions $V$ and $VI$ in the $(u, v)$ plane. As a result, the complete geodesics in $(t, r)$, described in the previous section, now correspond to continuous flows of information across the singularity in $(u, v)$. The singularity at $uv = 1$ is traversed smoothly, at least by massless geodesics displayed in Eqs.(44,45), rendering the spacetime geodesically complete.

The construction of the $(u, v)$ spacetime from $(r, t)$ proceeds as follows. Define the tortoise coordinate $r_*(r)$ through $(A(r))^{-1} dr = dr_*$ (see Eq.(25.31) in reference [14]). The metric Eq.(28) can then be written in $(t, r_*)$ as

$$ds^2 = A(r)(-dt^2 + dr_*^2) + r^2 d\Omega^2.$$

The KS coordinates are defined by

$$uv = \left(e^{r_*(r)/r_0}\right)^{\text{Sign}(r)} \times \text{Sign}\left(-rA\left(r\right)\right), \quad \frac{v}{u} = e^{t/r_0} \times \text{Sign}\left(-rA\left(r\right)\right),$$
$$\text{where } r_*(r) = \int_0^r (A(r'))^{-1} dr', \text{ with } -\infty < r < \infty. \tag{47}$$

The insertions of $\text{Sign}(r)$ and $\text{Sign}(-rA(r))$, for general $A(r)$, are novel in the literature and consistent with the construction in [9]. They ensure continuity of the $(r, t)$ to $(u, v)$ map. The sign functions can be expressed equivalently in terms of $r$ or the product $uv$:

$$\text{Sign}\left(-rA\left(r\right)\right) = \text{Sign}\left(uv\right). \tag{48}$$

From Eq.(47) one derives:

$$v^2 = e^{t/r_0} \left(e^{r_*(r)/r_0}\right)^{\text{Sign}(r)}, \quad u^2 = e^{-t/r_0} \left(e^{r_*(r)/r_0}\right)^{\text{Sign}(r)}$$
$$2\frac{dv}{v} = \frac{dt + \text{Sign}(r) dr_*}{r_0}, \quad 2\frac{du}{u} = \frac{-dt + \text{Sign}(r) dr_*}{r_0} \tag{49}$$
$$\frac{ds^2}{r_0^2} = \text{Sign}(r) |A(r)| \left(e^{-r_*(r)/r_0}\right)^{\text{Sign}(r)} (-4 du dv) + \frac{r^2}{r_0^2} d\Omega^2.$$

In the final expression, the $r$-dependent functions for arbitrary $A(r)$ can be rewritten purely in terms of the product $uv$ using Eqs.(47-50).

Once expressed in terms of only $uv$ and $dudv$ the metric in the maximally extended KS spacetime, i.e. $ds^2$ above, fully agrees with [9][6] in the limit of zero curvature $R_\pm \to 0$.

The tortoise coordinate $(r)$ and the function $uv(r)$ in Eq.(47) can be evaluated analytically for the $A^{-1}(r)$ given in (34):

$$r_*(r) = \begin{cases} \theta(r) \left[\rho_h \ln\left|1 - \frac{r}{r_h}\right| + \rho_c \ln\left|1 - \frac{r}{r_c}\right| - (\rho_h + \rho_c) \ln\left(1 + \frac{r}{r_h + r_c}\right)\right] \\ +\theta(-r) \left[\rho_w \ln\left(1 - \frac{r}{w}\right) + \rho_w^* \ln\left(1 - \frac{r}{w^*}\right) - (\rho_w + \rho_w^*) \ln\left(1 + \frac{r}{w + w^*}\right)\right] \end{cases},$$

$$uv(r) = \begin{cases} \theta(r) \left|1 - \frac{r}{r_h}\right|^{\rho_h/r_0} \left|1 - \frac{r}{r_c}\right|^{\rho_c/r_0} \left(1 + \frac{r}{r_h + r_c}\right)^{-(\rho_h + \rho_c)/r_0} \text{Sign}\left(\left(1 - \frac{r}{r_h}\right)\left(1 - \frac{r}{r_c}\right)\right) \\ +\theta(-r) \left(1 - \frac{r}{w}\right)^{-\rho_w/r_0} \left(1 - \frac{r}{w^*}\right)^{-\rho_w^*/r_0} \left(1 + \frac{r}{w + w^*}\right)^{(\rho_w + \rho_w^*)/r_0} \end{cases}.$$

$$\tag{50}$$

---

[6] In comparing expressions as functions of $(t, r)$ in this paper versus [9], the reader must take into account that the meaning of the symbol $r$ is different. Namely, the treatment in [9] kept the symbol $r$ strictly positive while allowing an antigravity domain of zero curvature.

Here $r_0$ can be expressed in terms of $(r_h, r_c)$ or $(w, w^*)$ as in Eq.(37), with the latter linked to $\beta$ through Eq.(37). The functions $r_*(r)$ and $uv(r)$ are plotted in Figs.(6,7,8), using illustrative unrealistic numerical values for $(r_h, r_c, \beta)$ to highlight the physical features. The tortoise coordinate $r_*$ is the area under the curve $(A(r))^{-1}$ in Fig.2, as measured from $r = 0$. This explains the behavior of $r_*(r)$ including the spikes at $r_h$ and $r_c$ which diverge to $\mp\infty$ respectively. At the singularity $r = 0$, where the gravity/antigravity transition occurs, both $r_*(r)$ and $uv(r)$ remain continuous. At $r = r_h$, the function $uv(r)$ exhibits a saddle-point-like behavior: both $uv(r)$ and its derivative $\partial_r(uv(r))$ vanish, while $uv(r)$ changes sign. A similar effect occurs at the cosmological horizon $r = r_c$, where $uv(r)$ and its derivative $\partial_r(uv(r))$ blow up while $uv(r)$ changes sign again.

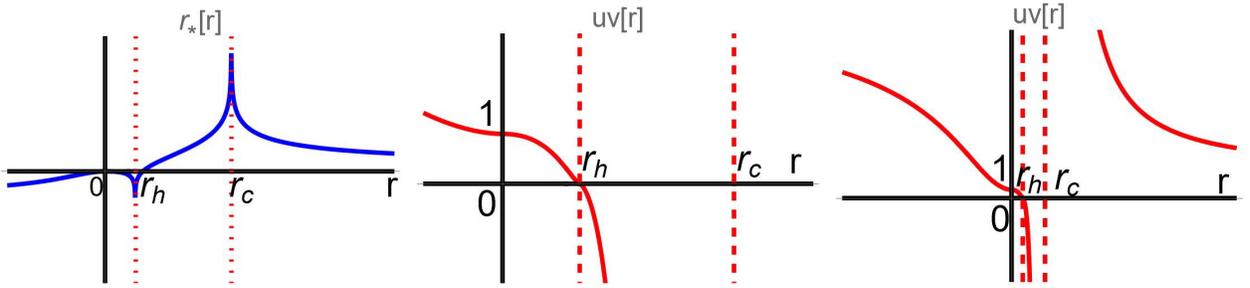

Fig.6 - $r_*(r)$ global behavior. Fig.7- $uv(r)$ near $r = 0$ & $r_h$. Fig.8- $uv(r)$ global behavior.

The behavior of Figs.7&8 can be quantified by expanding $uv(r)$ from Eq.(50) near the critical points $r = 0, r_h, r_c$, and $\pm\infty$:

| limits | $uv(r)$ | Computation |
|---|---|---|
| $r \to +\infty$ | $uv_{+\infty} \simeq$ | $\frac{r_c}{r_h \sqrt{e}} \simeq 1.4 \times 10^{17} \times \frac{r_\odot}{r_0}$ |
| $r \to r_c$ | $uv_c =$ | $\pm\infty$ |
| $r \to r_h$ | $uv_h =$ | $0$ |
| $r \to 0$ | $uv_0 =$ | $1$ |
| $r \to -\infty$ | $uv_{-\infty} \simeq$ | $\begin{cases} 1 & \text{for } \beta \approx 0^+, \\ e^{\frac{\pi/2}{\sqrt{1-\beta}}} \sqrt{e(1-\beta)} & \text{for } \beta \approx 1^-. \end{cases}$ |

(51)

The limiting values for $uv$ at $r \to r_0, r_h, r_c$ given in Eq.(51) are confirmed visually in Figs.7&8. The $r \to +\infty$ entry in Eq.(51) gives an approximation for $uv_{+\infty}$ in the realistic regime when $r_c/r_h$ is large, as in our own universe (see (32)). The $r \to -\infty$ entry is for the limiting cases for $uv_{-\infty}(\beta)$, evaluated in the $\beta \to 0$ or $1$ limits where $0 < \beta < 1$ remains a phenomenologically undetermined parameter. Thus $uv_{-\infty}$ is slightly above 1 in the weak antigravity limit ($\beta \approx 0$, $\tilde{G}_N \approx 0$, $R_- \approx -\infty$) and grows very large in the strong antigravity limit ($\beta \approx 1$, $\tilde{G}_N \approx \infty$, $R_- \approx 0$).

The exact asymptotic values $uv_{\pm\infty}$ are obtained directly from Eq.(50)

$$uv_{+\infty} = \left(1 + \frac{r_c}{r_h}\right)^{\rho_h/r_0} \left(1 + \frac{r_h}{r_c}\right)^{\rho_c/r_0},$$

$$uv_{-\infty} = \begin{cases} \left(\frac{1}{4} + \left(\frac{\operatorname{Im}(w)}{2\operatorname{Re}(w)}\right)^2\right)^{\operatorname{Re}(\rho_w/r_0)} e^{-2\left(\tan^{-1}\left|\frac{\operatorname{Im}(w)}{\operatorname{Re}(w)}\right| - \pi\right) \operatorname{Im}(\rho_w/r_0)} \\ = \left((1-\beta)\exp\left[-2\left(\tan^{-1}\sqrt{\frac{3+\beta}{1-\beta}} - \pi\right)\sqrt{\frac{3+\beta}{1-\beta}}\frac{3-\beta}{3+\beta}\right]\right)^{\frac{\beta^2}{6-4\beta}} \end{cases} . \quad (52)$$

Consistent with Figs.7&8, one finds $uv_{+\infty} > 1$ for any $(r_h/r_c) < 1$ and $uv_{-\infty} > 1$ for any $0 < \beta < 1$.

## VI. KRUSKAL-SZEKERES AND PENROSE DIAGRAMS

The Anti–de Sitter–Schwarzschild–de Sitter (AdSSdS) black hole spacetime in $(t, r)$ coordinates, Eq.(28), was extended in the previous section to a geodesically complete spacetime in Kruskal–Szekeres (KS) coordinates $(u, v)$, Eqs.(47-52). The plot of $uv(r)$ in Figs.7&8 is an essential tool that permits the construction and interpretation of the KS diagrams in Figs.9&10 and the Penrose diagram in Fig.11. I will refer to this larger extended structure simply as the AdSSdS black hole.

The KS diagram in Fig.9 is centered around the point at the origin $(u, v) = (0, 0)$, while the one in Fig.10 is centered around the point at infinity $(u, v) = (\pm\infty, \pm\infty)$. The switch of sign $\pm\infty$, that occurs at the cosmological horizon is understood through Fig.8. The $r > 0$ gravity regions that are familiar in the literature of the traditional KS diagrams are all the white regions in Figs.9&10. The new $r < 0$ antigravity regions in the AdSSdS black hole are painted yellow in Fig.9. The gray region in Fig.9 that satisfies $uv > uv_{-\infty} > 1$, and the blue region in Fig.10 that satisfies $uv < uv_{+\infty}$, are discarded because they are incompatible with the possible values of $uv(r)$ for all $-\infty < r < +\infty$ as illustrated in Fig.7&8.

Recall that in the vanishing antigravity curvature limit $R_- \to 0$ ($\beta \to 1$, strong antigravity), $uv_{-\infty}$ becomes infinitely large (Eq.(51)), so the gray region in Fig.9 disappears as the yellow region grows to fill the entire region $V_h$. In reverse, as the antigravity curvature grows in magnitude, $R_- \to -\infty$ ($\beta \to 0$, weak antigravity), $uv_{-\infty} \to 1$, so the gray region in Fig.9 fills the entire region $V_h$ as the yellow region shrinks to zero.

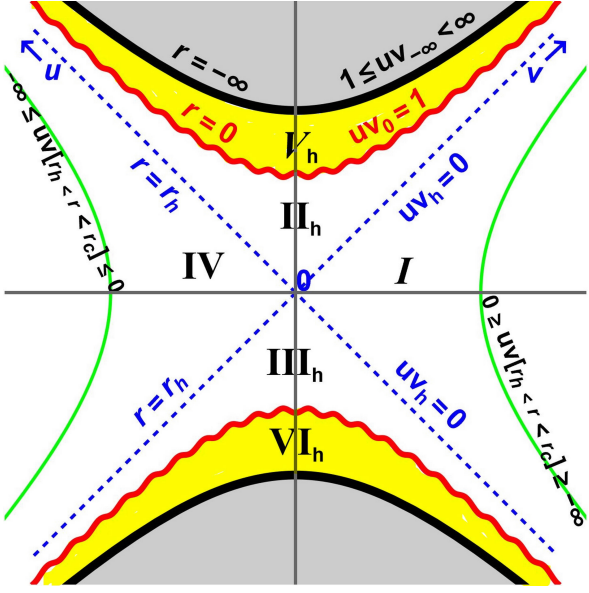 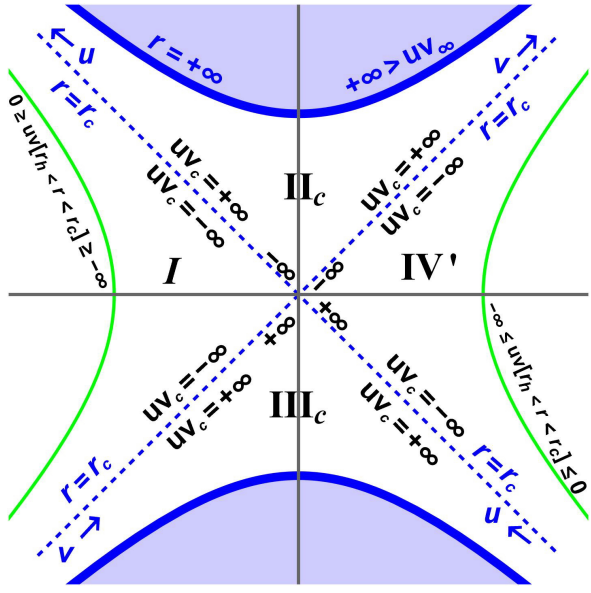

Fig.9- $(u,v)$ near black hole horizon.    Fig.10- $(u,v)$ near cosmological horizon.

Besides these remarks, the features of the KS diagrams are standard. Concentrating first on the white gravity regions, $(I, II_h, III_h, IV)$ in Fig.9 and $(I, IV')$ in Fig.10, from Figs.7&8 one can see that they correspond to $0 < r < r_c$ or $-\infty < uv < 1$, while the white regions $(II_c, III_c)$ in Fig.10 correspond to $r_c < r < +\infty$ or $uv_\infty < uv(r) < uv_c = +\infty$. One boundary of the white gravity region is the black hole singularity represented in Fig.9 by the red wavy hyperbolas that satisfy $uv = 1$ (at $r = 0$, see Fig.7); the second boundary at the edges of regions $II_c$ and $III_c$ in Fig.10, are represented by the blue hyperbolas that satisfy $uv = uv_\infty$ (at $r = +\infty$ see Fig.8). The green hyperbolas in both Figs.9&10 correspond to any finite value of $r$ in the visible region $r_h < r < r_c$ where $uv(r)$ is negative as seen in Figs.7&8.

The new yellow antigravity regions $V_h$ and $VI_h$ in Fig.9 are bounded by the red wavy hyperbolas that satisfy $uv = 1$ (at $r = 0$, see Fig.7) and the black hyperbola at the edge of $V_h$ that satisfies $uv = uv_{-\infty}$ (at $r = -\infty$, see Fig.7).

The blue dashed diagonal lines in Fig.9 correspond to the black hole horizon where $uv(r_h) = 0$ as seen in Fig.7, so either $u = 0$ at any $v$, or $v = 0$ at any $u$. Similarly, in Fig.10 the dashed lines correspond to the cosmological horizon where $uv(r_c) = \pm\infty$ as seen in Fig.8. Hence region $I$ is the visible region outside of the black hole, while $IV$ and $IV'$ are infinite mirror regions behind the two horizons.

The Penrose diagram in Fig. 11, in $(\tilde{u}, \tilde{v})$ coordinates, provides a clearer view of geodesic completeness and causality. The Penrose $(\tilde{u}, \tilde{v})$ coordinates are related to the KS coordinates $(u, v)$ via the conformal transformation

$$u = \tan \tilde{u}, \quad v = \tan \tilde{v}. \tag{53}$$

This mapping sends the 16 infinite regions, in the pair of KS diagrams in Figs.9&10, into 16 finite regions in the $(\tilde{u}, \tilde{v})$ plane, labeled by the same symbols, as shown in Fig. 11. Moreover, Fig. 11 represents only part of the full Penrose diagram. The latter is obtained by periodically repeating Fig.11 to the left and right ad infinitum. This follows from the $\pm 2\pi n$ periodicity of the tangent function in (53).

The product $uv$, evaluated at the critical values in Figs.7&8, becomes

$$uv\left(r\right) \Rightarrow \tan \tilde{u}\left(r\right) \tan \tilde{v}\left(r\right)|_{r\to(-\infty, 0, r_h, r_c, +\infty)} = \left(uv_{-\infty}, 1, 0, \pm\infty, uv_{+\infty}\right). \tag{54}$$

These equations define curve segments in the $(\tilde{u}, \tilde{v})$ plane, forming the region boundaries in Fig. 11. For example, at the $r = 0$ singularity, $\tan \tilde{u}(0) \tan \tilde{v}(0) = 1$, can be rewritten as $\cos(\tilde{u}(0) + \tilde{v}(0)) = 0$, which implies $\tilde{u}(0) + \tilde{v}(0) = \pm\pi/2 \pm 2\pi n$. This gives the red horizontal wavy line segments in the infinitely periodic version of Fig.11[7]. At the other critical points in (54), the trigonometry algebra does not yield simple expressions, but a plot can be obtained easily, showing that Eq.(54) produce the region boundaries in the infinitely periodic version of Fig.11.

Therefore, the region boundaries in Fig.11 are interpreted as follows.

- The blue convex curves at top $I^+$ and bottom $I^-$, represent the asymptotic boundary of the dS gravity region at $r = +\infty$.

---

[7] This is more clearly seen by a change of variables $(\tilde{u}, \tilde{v})$ to $(\tilde{t}, \tilde{r})$ by: $\tilde{u} = (\tilde{t} - \tilde{r})$ and $\tilde{v} = (\tilde{t} + \tilde{r})$. Here space $\tilde{r}$ runs horizontally, and time $\tilde{t}$ runs vertically, and the origin $\tilde{t} = \tilde{r} = 0$ is at the center of Fig.11. Then the segment equation can be rewritten as $\cos(\tilde{u}(0) + \tilde{v}(0)) = 0 = \cos 2\tilde{t}$. This is satified by $\tilde{t} = \pm\frac{\pi}{4}$ in the two intervals $-\frac{\pi}{4} < \tilde{r} < \frac{\pi}{4}$ and $\left(-\frac{\pi}{4} \pm \pi\right) < \tilde{r} < \left(\frac{\pi}{4} \pm \pi\right)$ that cover the mapping of Figs.9&10, as shown in Fig.11. In addition, the red wavy segments appear in the $2\pi$ periodic repetitions of $\tilde{r} \to \tilde{r} \pm 2\pi n$. This is why Fig.11 is periodically repeated indefinitely to the left and to the right. One may ask: why is not repeated periodically also in the $\tilde{t}$ direction? The answer is because complete geodesics must end at $r = \infty$ (not $\tilde{r}$). The $r = \infty$ boundary corresponds to the blue curves in $(\tilde{r}, \tilde{t})$ space as indicated in Fig.11 at the top (labelled as $\mathcal{I}^+$) and at the bottom (labelled as $\mathcal{I}^-$). This is in contrast to the black boundaries at the top and bottom labelled as $\mathcal{I}^\pm$ that correspond to $r = -\infty$, where geodesics don't end, but rather they get reflected as explained at the end of this section. Since geodesics cannot continue vertically in the perpendicular $\tilde{t}$ direction, it is of no use to consider the periodicity in $\tilde{t}$.

- The geen diagonal segments, represent the future and past cosmological horizons at $r = r_c$.

- The black & green diagonal dashed segments, represent the future and past black-hole horizons at $r = r_h$.

- The red wavy segments at top and bottom, represent the black- and white-hole singularities at $r = 0$.

- The black concave curves labelled $I^{\pm}$ at the top and bottom, represent the black- and white-hole internal asymptotic boundary of the AdS antigravity regions at $r = -\infty$.

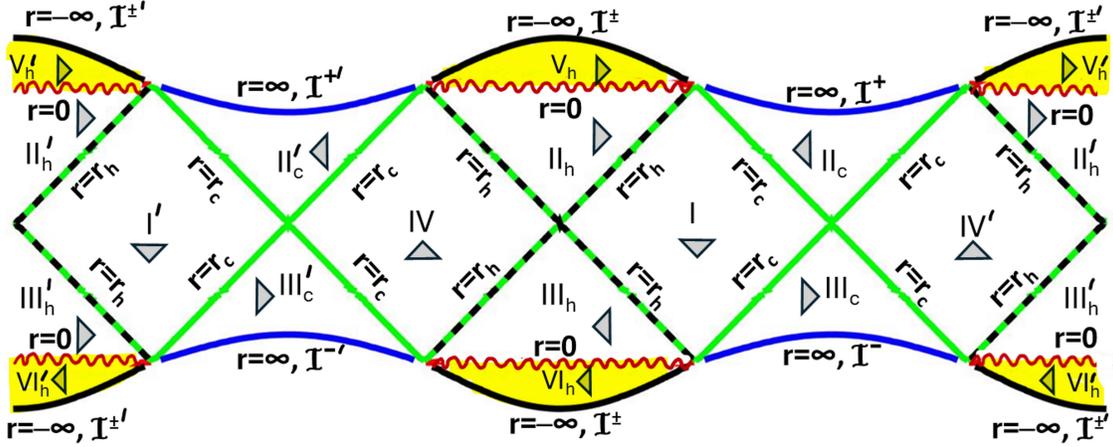

Fig.11 - Penrose diagram $(\tilde{u}, \tilde{v})$. The geodesically complete and causal spacetime.

The antigravity domains (yellow) lie between the red wavy segments and the black concave curves and are labelled $(V_h, V_h')$, $(VI_h, VI_h')$. All white regions correspond to the standard gravity domains $(I, II_h, III_h, IV)$ and $(I', II_c', III_c', IV')$, and they resemble the Penrose diagram for the SdS black hole given by Gibbons and Hawking (see Fig.4 in [11]). The AdSSdS diagram in Fig. 5 is geodesically complete, unlike the incomplete SdS case in [11].

Small right-angled triangles throughout the diagram indicate local forward lightcones. A massive particle located at the right-angle vertex of a triangle must propagate within the forward cone; a massless particle travels only along one of the cone's right-angle edges. These causal rules follow from the Killing vector $\partial_t$ associated with the conserved energy $E$. All geodesics must obey these propagation rules at each instant of proper time.

*Example*: Consider a massless particle (e.g. photon) in region $I$ at radial position $r_0$ with $r_h < r_0 < r_c$, moving radially toward the black hole with vanishing angular momentum $\vec{L} = 0$. It's geodesic (see Eq.(45)) is a $(3\pi/4)°$ straight line parallel to the lower $r = r_h$ horizon. As proper time increases, the photon crosses into region $II_h$, passes through the upper $r = r_h$ horizon, and reaches the singularity $r = 0$ (red wavy line) in a finite amount of proper time, $\tau_0 = \frac{r_0}{E}$, where $E$ is the photon energy. The geodesic then continues through the singularity into antigravity region $V_h$, requiring an *infinite* amount of proper time $\tau$ to reach the $r = -\infty$ boundary (black curve). This appears to be a complete geodesic because it is not artificially truncated at a finite value of proper time, as it would have happened if the antigravity region had been excised (as in [11]).

Actually, the geodesic above is not complete yet in the AdSSdS geometry of Fig.5. Besides the infinite range of proper time $\tau$ which must not be cutoff artificially, one must also consider what is called the global geometry[8] of the AdS spacetime [16]. According to the global geometry, AdS spacetime is analogous to a box, such that the $r = -\infty$ boundary marked as $\mathcal{I}^{\pm}$ in Fig.5. acts like a mirror at the end of the box. So although, $\mathcal{I}^{\pm}$ at $r = -\infty$, is reached in infinite amount of proper time, it is not the end of global conformal time (this can also be gathered from Fig.6 by noting that $t \sim -r_*(r)$ in Eq.(45) is finite as $r \to -\infty$). Accordingly, the geodesic discussed above does not end there, it is reflected at the $\mathcal{I}^{\pm}$ boundaries, and then moves downward at an angle of $(\pi/4)°$ following the only permitted causal lightcone direction (the little triangle) in region $V_h$. So, it continues as a $(\pi/4)°$ straight line that goes through the antigravity region $V_h$, sails through the singularity (wavy line), then the gravity regions $II_h$, then $IV$, and so on.

Similarly, one can easily figure out the complete geodesics of massless particles that originate in any region of the AdSSdS Penrose diagram in Fig.5. This overall picture of complete geodesics alters radically the discussion concerning the information paradox in black holes: see next section.

---

[8] From the perspective of 2T-physics, that predicts the conformal properties if i(SM+GR), global coordinates must be used to cover the entire 1T-spacetime, dS or AdS, both as derived by gauge fixing global flat 4+2 dimensional 2T-spacetime [xxx].[22]

## VII.  GLOBAL UNIFIED GRAVITY–ANTIGRAVITY SPACETIME

For clarity and tractability, the present work has modeled black holes as *eternal* black holes. Their Penrose diagrams are therefore drawn in their maximally extended form, without truncation to account for specific formation mechanisms such as primordial generation, stellar collapse, or binary mergers. Each of these scenarios certainly leaves distinctive imprints on the detailed causal structure, and incorporating them is an important direction for future refinements. Nevertheless, they are set aside here in order to highlight the fundamental global properties of spacetime itself, independent of astrophysical contingencies. By working in this idealized setting we expose the essential role of geodesic completeness and the new domains predicted by $i(\text{SM} + \text{GR})$.

Up to this point, the focus has been on the geometry associated with a *single* black hole. However, astrophysical observations show that the universe contains an enormous population of black holes, ranging from stellar-mass remnants to supermassive objects at galactic centers. According to $i(\text{SM}+\text{GR})$, each of these black holes contains, deep within, an *antigravity domain*. These interior regions are an intrinsic prediction of the local scale-symmetric formulation, and they represent spacetime sectors entirely absent in the conventional SM+GR framework. A natural question immediately arises: are these hidden antigravity interiors mutually connected, forming a vast web of causally linked regions, or do they remain isolated, each sealed within the event horizons of its parent black hole? The answer is not yet known, but either possibility has profound implications. What is certain, however, is that a truly global, geodesically complete universe must take into account not only the traditional gravitational regions both outside and inside black holes, but also *all* of the antigravity interiors (yellow in Fig. 5). Moreover, this global structure necessarily includes the additional regions beyond cosmological horizons—denoted $II_c, III_c, II'_c, III'_c$—together with their infinite, periodic extensions indicated schematically in Fig. 5.

Within such a geodesically complete framework, the celebrated *information puzzle* takes on a different character. In the conventional SM+GR description, and likewise in most current quantum gravity proposals, the antigravity regions do not appear at all. Their absence leaves the global picture incomplete and ensures that trajectories of infalling matter terminate at singularities without a consistent continuation. This truncation is precisely what perpetuates the information paradox: if the geodesics are artificially ended, information seemingly disappears. By contrast, $i(\text{SM} + \text{GR})$ offers a natural resolution. Here, classical information carried by infalling matter is not lost to all observers. Yes, it is lost partially for observers outside of black holes (except for comes back quantum mechanically), but what is lost to them is gained by observers in the interior regions, including antigravity domains. Hence information is transferred—at least in part—through massless carriers such as gravitons, photons, and gluons into definite regions of the full, extended spacetime. Thus, within this framework, the question "*Where does the information go?*" has a well-defined classical physics answer: it propagates into regions that were previously unknown in incomplete models. In quantum physics, we need to also include quantum tunnelling as well as quantum reflection at the effective potential barriers located at the singularity.

This perspective also dovetails naturally with extensions toward string theory. The reasoning outlined here at the level of classical field theory admits a parallel generalization to string-theoretic models, in line with approaches suggested in [5]. If developed further, this may open entirely new directions for understanding the deeper quantum structure of spacetime.

A further conceptual advance concerns the issue of *unitarity*. In ordinary discussions, unitarity is framed with respect to observers restricted to region $I$, namely the visible external universe. Information that crosses the horizon and fails to reemerge appears to imply a breakdown of quantum unitarity. However, once the full $i(\text{SM} + \text{GR})$ geometry is acknowledged as being physical, this conclusion is no longer warranted. Unitarity must instead be defined with respect to the *complete* spacetime, including not only region $I$, but also the black hole interiors, the regions beyond horizons, and crucially, the antigravity sectors that required for geodesic completeness. From this broader perspective, information is never lost: what vanishes from one observer's causal domain simply reappears as accessible information to another observer in a different, geodesically connected domain. The flow of information across the global manifold, illustrated schematically in Fig. 5, ensures that overall conservation of information is respected.

This global viewpoint also intersects fruitfully with ideas from quantum information theory, especially conjectures such as ER = EPR [17][18]. The present work provides evidence, at the classical level, for communication channels connecting region $I$ with region $IV$, mediated through black hole and white hole structures (with the caveat that the latter must be physically realized). Such *classical paths of information transfer* were not accounted for in previous treatments of black hole information flow. Their existence points toward new mechanisms by which entanglement and connectivity across spacetime might be physically grounded.

The extensive framework of AdS/CFT [19]-[21] is also given new life in this context. In standard discussions, AdS geometries serve as toy models or holographic laboratories. In the present picture, however, an *actual AdS region* arises as a genuine component of the antigravity interior geometry of a black hole in i(SM+GR). This identification transforms what was once a mathematical convenience into a physical reality, suggesting that insights from holography may be directly applicable to the study of antigravity interiors.

Finally, it is natural to ask whether this unified gravity–antigravity framework makes contact with *observable physics*. The answer is affirmative. Once the profiles $\phi(r)$ and $s(r)$ are fully determined (as shown schematically in Fig. 1), the associated metric $g_{\mu\nu}(r)$ will necessarily deviate from the form given in Eq. (28). These deviations modify spacetime curvature not only inside the horizon but also outside it, where they can in principle be probed. Such modifications may influence astrophysical observables, for example by altering galactic rotation curves, gavitational lensing, or affecting the dynamics of stars and gas in the vicinity of black holes. With sufficiently precise measurements, these addional curvature-induced effects could be distinguished from, or act in tandem with, the influence attributed to dark matter. This opens a novel phenomenological pathway to test the predictions of $i(\text{SM} + \text{GR})$. The explicit computations of the coupled fields $(\phi(r), s(r), g_{\mu\nu}(r))$ and their astrophysical implications will be presented in detail in [8].

---